\documentclass[onecolumn,10pt,showpacs,showkeys,preprintnumbers,amsmath,amssymb,usenames]{revtex4}
\usepackage{graphicx}
\usepackage{epsfig}
\usepackage{xcolor}
\def\dd{\displaystyle}
\begin{document}
\title{\bf Radiative Correction to the Casimir Energy for Lorentz-violating Scalar Field in $d+1$ Dimensions}
\author{M. A. Valuyan}
\email{m-valuyan@sbu.ac.ir; m.valuyan@semnaniau.ac.ir}
\affiliation{Department of Physics, Semnan Branch, Islamic Azad University, Semnan, Iran}
\date{\today}
\begin{abstract}
The renormalization program in every renormalized theory should be run consistently with the type of boundary condition imposed on quantum fields. To maintain this consistency, the counterterms usually appear in the position-dependent form. In the present study, using such counterterms, we calculated the radiative correction to the Casimir energy for massive and massless Lorentz-violating scalar field constrained with Dirichlet boundary condition between two parallel plates in $d$ spatial dimensions. In the calculation procedure, to remove infinities appearing in the vacuum energies, the box subtraction scheme supplemented by the cutoff regularization technique and analytic continuation technique were employed. Normally, in the box subtraction scheme, two similar configurations are defined and their vacuum energies are subtracted from each other in the appropriate limits. Our final results regarding all spatial dimensions were convergent and consistent with the expected physical basis. We further plotted the Casimir energy density for the time-like and space-like Lorentz-violating systems in a number of odd and even dimensions; multiple aspects of the obtained results were ultimately discussed.
\end{abstract}

\maketitle

\section{Introduction}\label{sec:intro}
Recent works have focused on Lorentz-violating systems, and different aspects of this symmetry breaking in the quantum field theory and quantum gravity are exciting to physicists. \textcolor[rgb]{0,0,0}{Earlier proposals regarding Lorentz violation were presented on spontaneous Lorentz symmetry breaking\,\cite{Nambu.,Pavlopoulos.,kostelecky}.} Following these proposals, other techniques of Lorentz symmetry violation were further introduced. Some of the mechanisms of symmetry breaking are: space-time non-commutativity\,\cite{carroll,Anisimov,carlson,Hewett,Bertolami.1}, modifications of quantum gravity\,\cite{Alfaro.1,Alfaro.2}, and the variation of coupling constants\,\cite{Lehnert,Anchordoqui,Bertolami.2}. The experimental measurements performed to find the effective value of the Lorentz symmetry breaking on the system is cumbersome. We maintain that this effective value should be explored in the physical quantity. The Casimir effect is known as an important phenomenon associated with the quantum field theory; therefore, it can be a suitable choice for exploring the experimental effects of the Lorentz symmetry breaking. A great number of studies have considered the Casimir energy for multiple Lorentz-violating quantum fields. The primary works focused on the leading-order Casimir energy in the Lorentz symmetry breaking theory for real scalar field between two parallel plates were conducted in\,\cite{Frank,Kharlanov,silva}. Later, this quantity was computed for spinor field with MIT bag model\,\cite{fermionic}. Moreover, next to the leading order of the Casimir energy for a Lorentz-violating scalar field confined between two parallel plates was performed in three spatial dimensions\,\cite{reza.nuclear}. In this article, we generically studied the next to leading-order of the Casimir energy for Lorentz-violating and self-interacting scalar field theory between two parallel plates in $d$ spatial dimensions. 
The present research investigated the first-order radiative correction to the Casimir energy for massive and massless scalar field confined with Dirichlet boundary condition between two parallel plates.
\par
The pioneering work concerning radiative correction to the Casimir energy was conducted by Bordag et al. more than 30 years ago\,\cite{Bordag.et.al.}. Later, a large number of studies were carried out on this correction for various quantum fields and geometries\,\cite{other.RC.21,other.RC.22,other.RC.23,other.RC.24}. The main component of calculating the radiative correction to the Casimir energy is the renormalization program. In this regard, determining the appropriate counterterm has become a challenging task in calculating radiative correction to the Casimir energy\,\cite{1D-Reza}. Typically, the use of counterterms in the renormalization program is to eliminate divergences caused by the bare parameters of Lagrangian\,(\emph{e.g.}, the bare mass of the quantum field and bare coupling constant). \textcolor[rgb]{0.00,0.00,0.00}{In most of the previous works, the \emph{free counterterm}\,(used for Minkowski space) was applied in renormalization programs\,\cite{free.counterterms.1,free.counterterms.2,free.counterterms.3,free.counterterms.4,reza.nuclear}. 
Additionally, in some articles, the influence of imposed boundary conditions was added in counterterms \,\cite{Albuquerque}.} However, in practice, the free counterterm was used in the space between the two boundary conditions, and a different counterterm was employed on the boundary surface\,\cite{Fosco}. Following this historical process, the need to use a monotonous counterterm consistent with the imposed boundary conditions was considered as a significant issue in\,\cite{EUR-Reza,1D-Reza}. The counterterms that they introduced, unlike the \textit{free counterterms}, are usually position-dependent, allowing all the influences of the dominant boundary conditions or backgrounds to be reflected in the renormalization program. Their counterterm generates a self-consistent program for the renormalization of the bare parameters of the Lagrangian. One of the main superiorities of such ilk of counterterm over the \textit{free counterterm} is the ability to renormalize the bare parameters of the Lagrangian\,\cite{2D-Man}. This preponderance for the use of position-dependent counterterms was first illuminated in the calculation of radiative correction to the Casimir energy in two spatial dimensions\,\cite{2D-Man}. Radiative correction to the Casimir energy for massive scalar field confined between a pair of parallel plates in two spatial dimensions was reported divergent in\,\cite{cavalcanti.1,cavalcanti.2,cavalcanti.3}. On the other hand, by recalculating this quantity via the position-dependent counterterm, the answer was obtained convergent and consistent with all the expected physical basis\,\cite{2D-Man}. Use of the position-dependent counterterms in the renormalization program even for problems defined in the curved space was also successful\,\cite{BSS.Curved.1,BSS.Curved.2}. In a part of the present article, the Lorentz symmetry was violated. We hold that this symmetry breaking should also alter the renormalization program, with all its influences reflected in the counterterms. Therefore, using free counterterms in the renormalization program may not be legitimate, hence the necessity of employing position-dependent counterterms is felt. This type of counterterm allows all the changes caused by the Lorentz symmetry breaking to be automatically imported in the renormalization program. Accordingly, we used the position-dependent counterterms in the renormalization program to calculate the radiative correction to the Casimir energy for Lorentz-violating scalar field between two parallel plates in all spatial dimensions. In any spatial dimension regarding both massive and massless cases, our general answer fulfills the necessary physical expectations in the appropriate limits.
\begin{figure}[th] \hspace{0cm}\includegraphics[width=7cm]{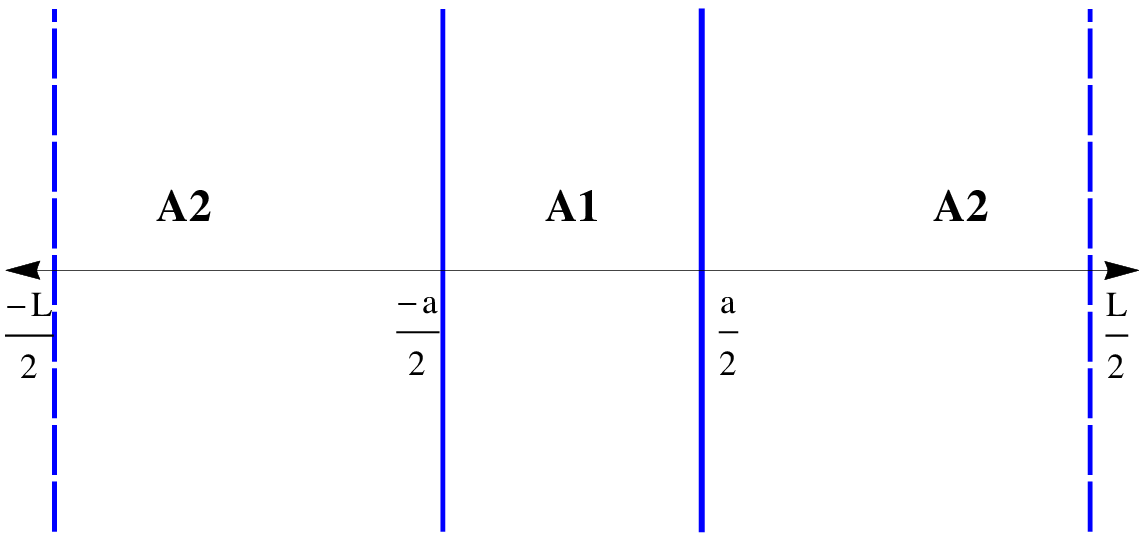}\hspace{1.3cm}\includegraphics[width=7cm]{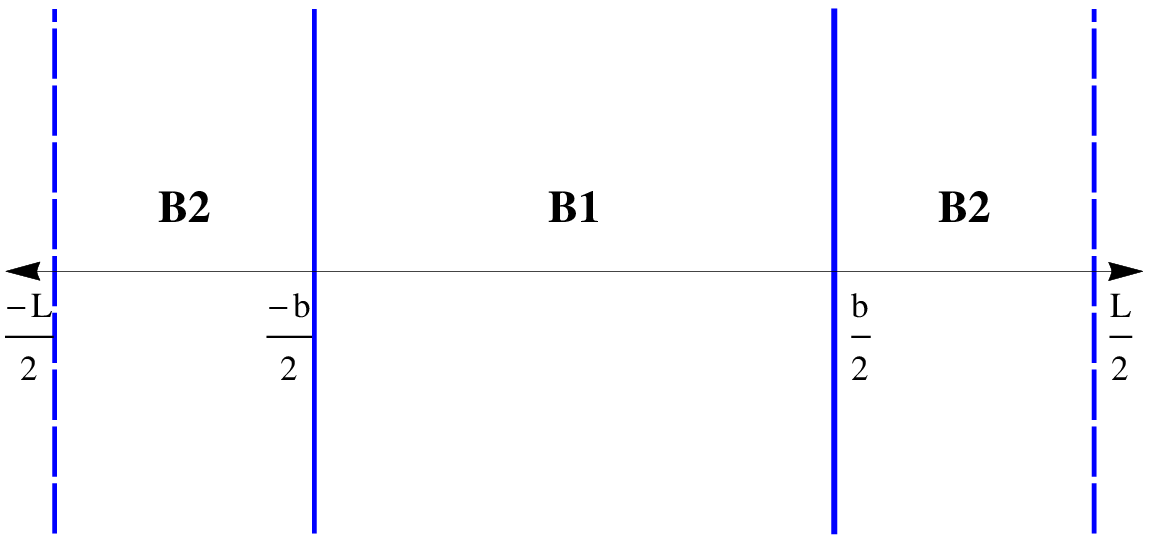}\caption{\label{BSS.Fig}
  The left figure is ``$A$ configuration" and the right one is ``$B$
  configuration".}
\end{figure}
\par
Dealing with divergent expressions makes an important part of calculating the Casimir energy. Therefore, to regularize the divergent expressions, different types of regularization technique were introduced\,\cite{phys.rep.,other.regular.tech.1,other.regular.tech.2,other.regular.tech.3,other.regular.tech.4,other.regular.tech.5,other.regular.tech.6,BSS.1,BSS.2,BSS.3}. In this study, to regularize and remove infinities caused by the vacuum energies, Box Subtraction Scheme\,(BSS), as a regularization technique were employed. Normally, two similar configurations are defined in the BSS. For instance, to calculate the Casimir energy between two parallel plates with a distance $a$, we need to place this pair of plates between two other parallel plates with a distance of $L>a$. The two outer plates play the role of a box for the two inner ones. In fig.\,(\ref{BSS.Fig}), we named this configuration as ``configuration $\mathcal{A}$''. Similar to configuration $\mathcal{A}$, we introduced another configuration, called $\mathcal{B}$. In configuration $\mathcal{B}$, two plates with distance $b>a$ were placed inside the two other plates with distance $L>b$. Now, we can define the Casimir energy by subtracting the vacuum energy of these two configurations according to the following expression:
\begin{eqnarray}\label{BSS.Def.}
  E_{\hbox{\tiny{Cas.}}}=\lim_{b/a\to\infty}\lim_{L/b\to\infty}[E_\mathcal{A}-E_\mathcal{B}],
\end{eqnarray}
where $E_{\mathcal{A}}$ and $E_{\mathcal{B}}$ are the vacuum energy of configurations $\mathcal{A}$ and $\mathcal{B}$, respectively. This definition of the Casimir energy is based on Boyer's method\,\cite{Boyer.}. In this paper, this method was generalized to higher dimensions to regularize and eliminate the divergences in the calculation process  related to the radiative correction to the Casimir energy. The paper is organized as follows: in the next section, we primarily discuss the summary of the renormalization program and the deduction of position-dependent counterterms for Lorentz-violating $\phi^4$ theory. The general form of the first-order vacuum energy expression was then calculated through  employing the obtained counterterms from the renormalization program. In Section \ref{sec: RC.calculation}, we computed the radiative correction to the Casimir energy for the massive and massless scalar field confined with the Dirichlet boundary condition between two parallel plates in $d$ spatial dimensions. Considering the conservation and breaking of the Lorentz symmetry, the radiative correction to the Casimir energy for both cases were obtained. Finally, in Section\,\ref{sec:conclusion}, all the obtained results and their related aspects were concluded.
\section{The Lorentz-violating $\phi^4$ Theory}\label{sec:The.Lorentz.violating.Model}
The Klein-Gordon Lagrangian with Lorentz-violating term is normally defined as\,\cite{cruz}:
\begin{eqnarray}\label{lagrangian.free.field}
   \mathcal{L}=\frac{1}{2}\big(\partial_\mu \phi \partial^\mu \phi+\beta(u\cdot\partial\phi)^2-m_0^2\phi^2\big),
\end{eqnarray}
where the parameter $m_0$ is the bare mass of the real scalar field, and the dimensionless coefficient $\beta$ shows the scale of the Lorentz symmetry breaking. This parameter is usually set to much smaller than one, and it is able to codify the Lorentz-violating through multiplying the derivative of the scalar field by a constant vector $u^\mu$. By changing the vector $u^\mu$, the direction of the Lorentz-violating can be oriented\,\cite{Colladay,Gomes}. The equation of motion related to the Lagrangian shown in eq.\,\eqref{lagrangian.free.field} reads as:
\begin{eqnarray}\label{KG.equation.of.motion}
   [ \Box+\beta(u.\partial)^2-m_0^2]\phi=0.
\end{eqnarray}
To present the radiative correction level to the Casimir energy, a theoretical model for self-interacting and Lorentz-violating massive scalar field is required. Therefore, we added a self-interacting term to the defined Lagrangian in eq.\,\eqref{lagrangian.free.field}. We obtain,
\begin{eqnarray}\label{lagrangian.bare}
   \mathcal{L}=\frac{1}{2}\big(\partial_\mu \phi \partial^\mu \phi+\beta(u\cdot\partial\phi)^2-m_0^2\phi^2\big)-\frac{\lambda_0}{4!}\phi^4,
\end{eqnarray}
where $\lambda_0$ is the bare coupling constant. At the level of the radiative correction to the Casimir energy, all bare parameters of Lagrangian\,(\emph{e.g.} $m_0$ and $\lambda_0$) must be renormalized. For this purpose, the scalar field is usually re-scaled by a field strength renoramlization parameter $Z$. After this re-scaling the Lagrangian is converted to,
\begin{eqnarray}\label{lagrangian.renormalized}
     \mathcal{L}=\frac{1}{2}(\partial_\mu\phi_r)^2+\frac{1}{2}\beta(u\cdot\partial\phi_r)^2-\frac{1}{2}m^2\phi_r^2-\frac{\lambda}{4!}\phi_r^4\nonumber\\
     +\frac{1}{2}\delta_Z(\partial_\mu\phi_r)^2+\frac{1}{2}\delta_Z\beta(u\cdot\partial\phi_r)^2-\frac{1}{2}\delta_m\phi^2-\frac{\delta_{\lambda}}{4!}\phi_r^4,
\end{eqnarray}
where $\phi=Z^{\frac{1}{2}}\phi_r$, $\delta_Z=Z-1$, $\delta_m=m_0^2Z-m^2$, and $\delta_\lambda=\lambda_0Z^2-\lambda$. Moreover, $m$ and $\lambda$ are the physical mass of the field and coupling constant, respectively. The Feynman rules associated with counterterms in the above Lagrangian are:
\begin{eqnarray}\label{feynman.rule.counterterm}
   \raisebox{0mm}{\includegraphics[width=1cm]{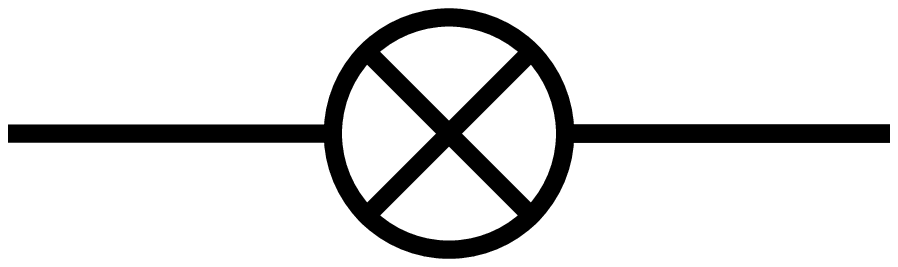}}&=&i\big[(p^2+\beta(u\cdot p)^2)\delta_Z-\delta_m\big],\nonumber\\
   \raisebox{-2mm}{\includegraphics[width=0.7cm]{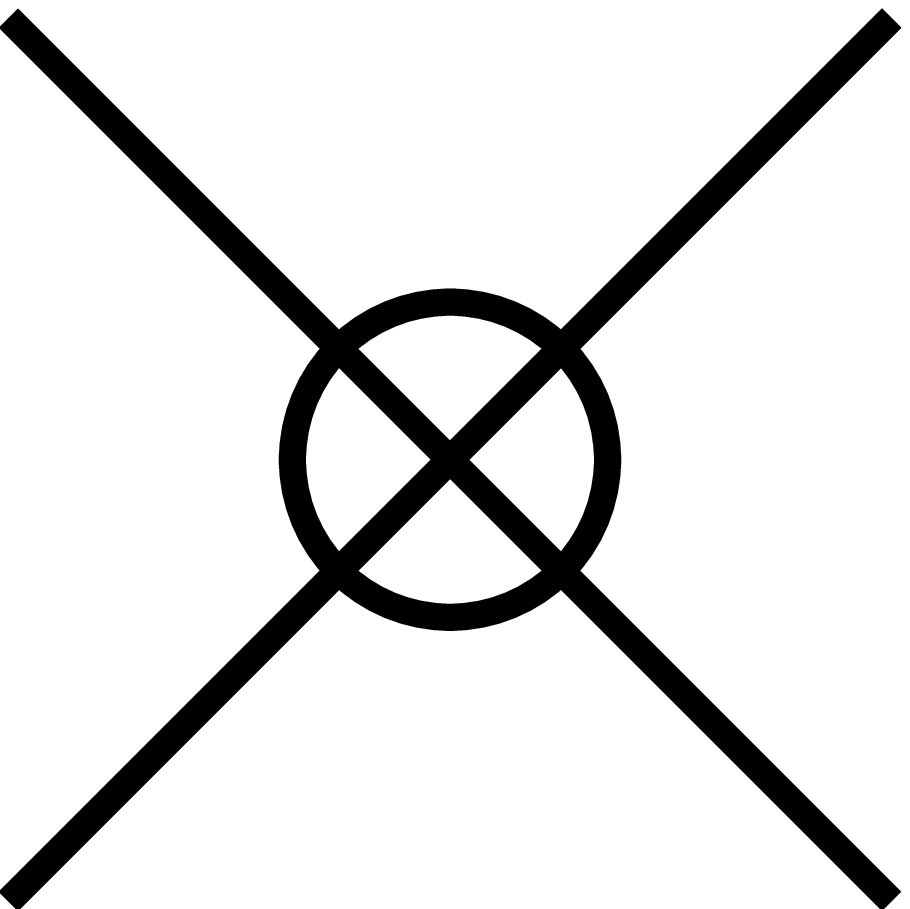}}&=&-i\delta_\lambda.
\end{eqnarray}
To determine the values of the counterterms, the following form of renormalization conditions should be applied:
\begin{eqnarray}\label{renormalization.conditions}
     \raisebox{-3mm}{\includegraphics[width=1cm]{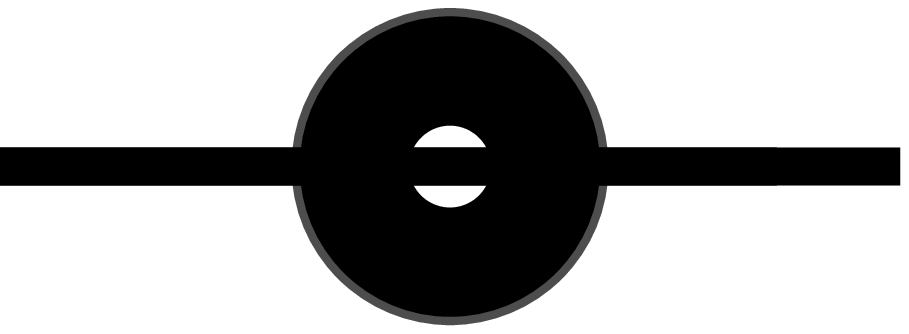}}&=&\frac{i}{p^2-m^2}+\mbox{(the terms regular at $p^2=m^2$)},\nonumber\\
   \raisebox{-2mm}{\includegraphics[width=0.7cm]{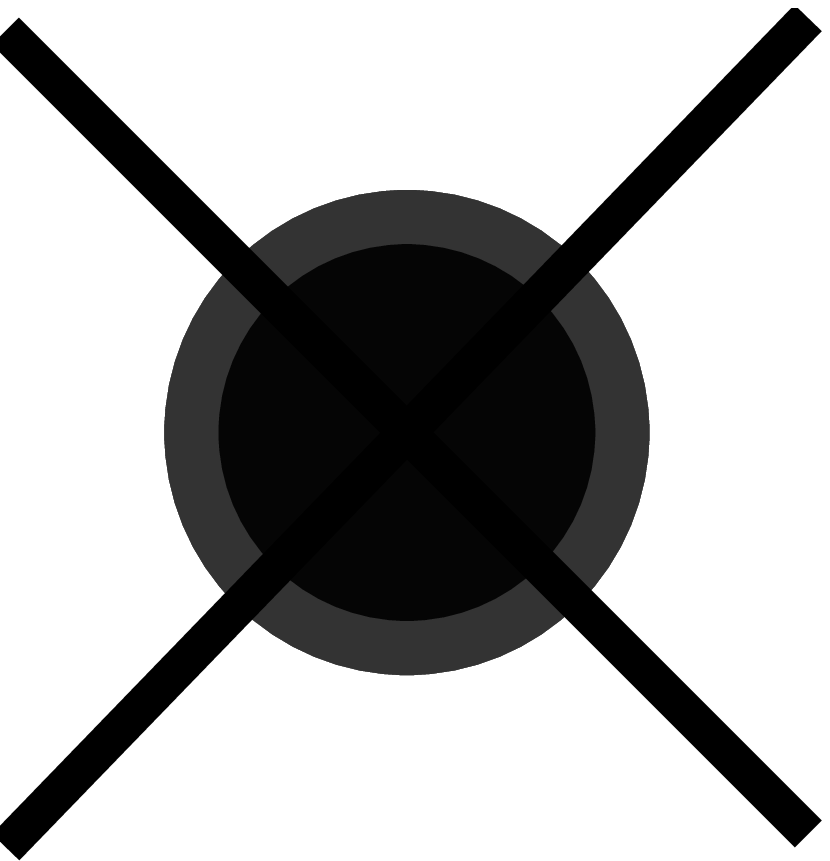}}&=&-i\lambda\hspace{2cm} \mbox{(at $s=4m^2$,$t=u=0$)}.
\end{eqnarray}
\textcolor[rgb]{0.00,0.00,0.00}{where $s$, $t$ and $u$ indicate the type of the channel. As known, the channel can be read from the form of the Feynman diagram, and each channel leads to characteristic angular dependence of the cross section.} The perturbative expansion pertaining to the two point function up to the first-order of the coupling constant $\lambda$ is usually written as:
\begin{eqnarray}\label{two.point.fuction}
      \raisebox{-2.7mm}{\includegraphics[width=1.2cm]{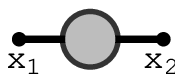}}= \raisebox{-2mm}{\includegraphics[width=1.2cm]{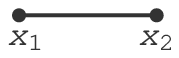}}+ \raisebox{-2.2mm}{\includegraphics[width=1cm]{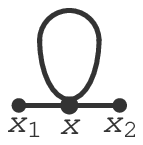}}+ \raisebox{-2.3mm}{\includegraphics[width=1.2cm]{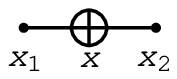}}.
\end{eqnarray}
Now, to fix the counterterms, the renormalization conditions should be applied to each order of coupling constant $\lambda$ in the perturbative expansion of the two point function displayed in eq.\,\eqref{two.point.fuction}. Doing so at the first order of coupling constant $\lambda$, the values of $\delta_\lambda$ and $\delta_Z$ become zero, and the expression for the mass counterterm is obtained as:
\begin{eqnarray}\label{counterterm.value}
      \delta_m(x)=\frac{-i}{2}\raisebox{-2.2mm}{\includegraphics[width=1cm]{15.eps}}=\frac{-\lambda}{2}G(x,x),
\end{eqnarray}
where $G(x,x)$ is the Green's function. Through eq.\,\eqref{counterterm.value} all effects of the boundary condition or non-trivial background, which we believe should influence the renormalization program, are reflected via the Green's function in the mass counterterm. By solving the motion equation given in eq.\,\eqref{KG.equation.of.motion} and applying the Dirichlet boundary condition to the quantum field at the plates placed on $z=\pm\frac{a}{2}$\,(region $A1$ of fig.\,(\ref{BSS.Fig})), when the Lorentz symmetry is still preserved ($\beta=0$), the following dispersion relation is obtained:
\begin{eqnarray}\label{dispersion.relation.DBC}
\omega_n^2=k_1^2+k_2^2+(\frac{n\pi}{a})^2+k_4^2+...+k_d^2+m_0^2,\hspace{1cm}n=1,2,3,...
\end{eqnarray}
where $k_3=k_z=\frac{n\pi}{a}$ is the wave vector perpendicular to the plates and $k_\perp=(k_1,k_2,k_4,...,k_d)$ denotes all other directions of the wave vector. The final expression of the Green's function for the real scalar field confined with Dirichlet boundary condition between two parallel plates with distance $a$ in arbitrary spatial dimension $d$ after Wick rotation becomes:
\begin{eqnarray}\label{Greens.Function}
    G(a;x,x')=\frac{2}{a}\int\frac{d^dk}{(2\pi)^d}\sum_{n=1}^{\infty}\frac{e^{-ik_{\perp}(x-x')}e^{-\omega(t-t')}
    \sin(\frac{n\pi}{a}(z+\frac{a}{2}))\sin(\frac{n\pi}{a}(z'+\frac{a}{2}))}
    {\omega^2+k_{\perp}^2+(\frac{n\pi}{a})^2+m^2},
\end{eqnarray}
where $k=(\omega,k_{\perp})$. The above Green's function expression was written for region $A1$ of fig.\,(\ref{BSS.Fig}) and the Lorentz symmetry was preserved\,($\beta=0$). In the Lorentz symmetry breaking, three general types of directions are possible in violating, namely time-like direction, a space-like direction parallel to the plates\,(\textit{i.e.}, boundary conditions), and a space-like perpendicular to the plates. Different vector $u^\mu$ allows the Lorentz symmetry to break at different directions. If vector $u^\mu$ is selected as $u^\mu=(1,0,0,...,0)$, the case of time-like\,(TL like) Lorentz-violating occur. In this case, after solving eq.\,\eqref{KG.equation.of.motion} and applying the Dirichlet boundary condition at the boundaries placed on $z=\pm a/2$ defined in fig.\,(\ref{BSS.Fig}), the dispersion relation becomes:
\begin{eqnarray}\label{dispersion.relation.TLDBC}
    (1+\beta)\omega_n^2=k_\perp^2+k_z^2+m_0^2.
\end{eqnarray}
Also, the Green's function expression in this case can be written as:
\begin{eqnarray}\label{Green.function.TLDBC}
    G_{TL}(a;x,x')=\frac{1}{(1+\beta)^{\frac{1}{2}}}G(a;x,x').
\end{eqnarray}
Admitting that vector $u^\mu$ is space-like, we will have $d$ different cases. In the number of $d-1$ cases, the vector $u^\mu$ is parallel to the plates and the dispersion relation in these $d-1$ cases is obtained as follows:
\begin{eqnarray}\label{dispersion.relation.SLPADBC}
    \omega_n^2=k_1^2+k_2^2+k_z^2+...+(1-\beta)k_i^2+....+k_d^2+m_0^2.
\end{eqnarray}
From here on, this type of Lorentz-violating direction is called ``SP-Par like''. Performing the usual process of calculation for the Green's function in the SP-Par like Lorentz-violating system, we obtain:
\begin{eqnarray}\label{Green.function.SLPADBC}
    G_{SP-Par}(a;x,x')=\frac{1}{(1-\beta)^{\frac{1}{2}}}G(a;x,x').
\end{eqnarray}
For the only remaining direction of $u^\mu=(0,0,0,1,0,...,0)$ that is perpendicular to the plates, after solving the motion equation displayed in eq.\,\eqref{KG.equation.of.motion}, we obtain the dispersion relation as:
\begin{eqnarray}\label{dispersion.relation.SLPRDBC}
    \omega_n^2=k_1^2+k_2^2+(\frac{n\pi}{\tilde{a}})^2+k_4^2+...+k_d^2+m_0^2,\hspace{2cm} \mbox{ $n=1,2,3,...$.}
\end{eqnarray}
where $\tilde{a}=\frac{a}{\sqrt{1-\beta}}$. For simplicity, we call this direction of Lorentz breaking ``SP-Perp'' like. In this case the Green's function expression becomes,
\begin{eqnarray}\label{Green.function.SLPRDBC}
 G_{SP-Perp}(a;x,x')=G(\tilde{a};x,x').
\end{eqnarray}
To achieve the radiative correction to the Casimir energy, the vacuum energy expression up to the first order of the coupling constat $\lambda$ is required. In the next section, by use of the Green's function given in eqs.\,\eqref{Greens.Function}, \,\eqref{Green.function.TLDBC}, \eqref{Green.function.SLPADBC}, and \eqref{Green.function.SLPRDBC}, this step of computation is followed for each case of Lorentz symmetry breaking.

\section{Radiative Correction to the Casimir Energy}\label{sec: RC.calculation}
In this section, we calculated the first-order radiative correction to the Casimir energy for the massive and massless scalar fields in $\phi^4$ theory between two parallel plates in $d+1$ dimensions. For this purpose, we start with the general form of the first-order vacuum energy as:
\begin{equation}\label{VacuumEn.damble.}
  E_{\mbox{\tiny Vac.}}^{(1)}=i\int d^d\mathbf{x}\bigg(\frac{1}{8}
  \raisebox{-7mm}{\includegraphics[width=0.5cm]{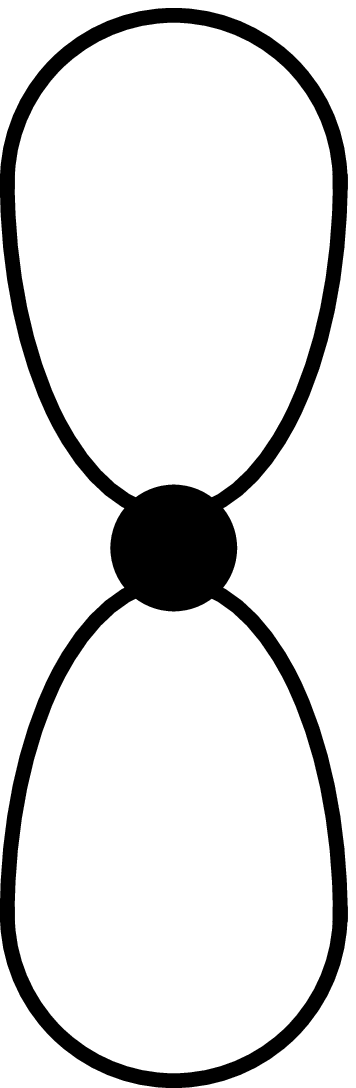}}+\frac{1}{2}\raisebox{-1mm}{\includegraphics[width=0.5cm]
  {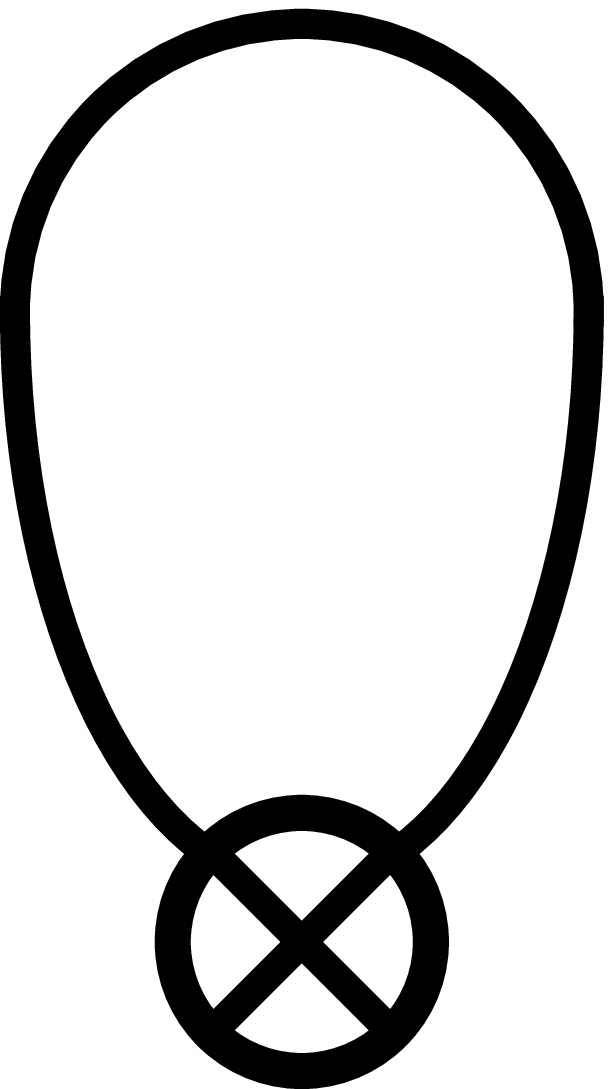}}\bigg)=i \int
  d^d\mathbf{x}\bigg(\frac{-i\lambda}{8}G^2(x,x)-\frac{i}{2}\delta_m(x)G(x,x)\bigg),
\end{equation}
where $G(x,x)$ is the Green's function, and the superscript $(1)$ denotes the first-order of this energy\,\cite{1D-Reza}. By substituting $\delta_m(x)$ from eq.\,\eqref{counterterm.value} in eq.\,\eqref{VacuumEn.damble.}, the total vacuum energy expression is obtained. Therefore, we have:
\begin{equation}\label{VacuumEn.firstorder.}
  E_{\mbox{\tiny Vac.}}^{(1)}=\frac{-\lambda}{8} \int G^2 (x,x)d^d\mathbf{x}.
\end{equation}
As shown in eq.\,\eqref{BSS.Def.}, to get the radiative correction to the Casimir energy, we need to obtain the vacuum energy of all regions pertaining to the two configurations displayed in fig.\,(\ref{BSS.Fig}). These energies should then be subtracted from each other in the appropriate limits. In the first onset, we present the details of this calculation for the system where the Lorentz symmetry is still preserved\,($\beta=0$). Afterwards, using the relations between the Green's function expressions displayed in eqs.\,\eqref{Green.function.TLDBC}, \eqref{Green.function.SLPADBC} and \eqref{Green.function.SLPRDBC}, we obtain the radiative correction to the Casimir energy for each case of the Lorentz-violating system. Accordingly, by substituting eq.\,\eqref{Greens.Function} with the vacuum energy given in eq.\,\eqref{VacuumEn.firstorder.} we obtain:
\begin{eqnarray}\label{After.Substituting.}
    E_{\mbox{\tiny Vac.}}^{(1)}(a)&=&\frac{-\lambda}{8} \int G^2 (a;x,x)d^d\mathbf{x}\nonumber\\&=&\frac{-\lambda}{8}\int_{-\frac{a}{2}}^{\frac{a}{2}}\frac{4}{a^2}\int_{0}^{\infty}\frac{\Omega_d t^{d-1}dt}{(2\pi)^d a^{d-2}}\sum_{n=1}^{\infty}\frac{\sin^2(\frac{n\pi}{a}(z+\frac{a}{2}))}{t^2+\omega_{a,n}^2}\int_{0}^{\infty}\frac{\Omega_d t'^{d-1}dt'}{(2\pi)^d
    a^{d-2}}
    \sum_{n'=1}^{\infty}\frac{\sin^2(\frac{n'\pi}{a}(z+\frac{a}{2}))}{t'^2+\omega_{a,n'}^2}dzL^{d-1},
\end{eqnarray}
where $\Omega_d=\frac{2\pi^{d/2}}{\Gamma(d/2)}$ is the spatial angle, $\omega_{a,n}^2=(n\pi)^2+(ma)^2$, and $t=ka$. To nondimensionalize the parameter $k$ in the integrand in eq.\,\eqref{After.Substituting.}, an appropriate factor $a$ was multiplied by the numerator and denominator of the integrand; next, the variable of the integrand was changed to $t=ka$. Calculating the integral over $t$ and $t'$ in eq.\,\eqref{After.Substituting.} and applying the analytic continuation technique for even values of dimension $d$, the vacuum energy of eq.\,\eqref{After.Substituting.} is converted to:
\begin{eqnarray}\label{computing.the.integral.A1.EVEN}
   E_{\mbox{\tiny Vac.}}^{(1)}(a)=\frac{-\lambda L^{d-1}\Omega_{d}^2}{8(2\pi)^{2d}a^{2d-3}}\sum_{n,n'=1}^{\infty}
                                        \big(1+\frac{1}{2}\delta_{n,n'}\big)\omega_{a,n}^{d-2}\omega_{n'}^{d-2}\ln\omega_{a,n}\ln\omega_{a,n'}, &\hspace{1.5cm}  d=2,4,6,8,....
\end{eqnarray}
Performing this procedure for odd values of dimension $d$ leads to:
\begin{eqnarray}\label{computing.the.integral.A1.ODD}
   E_{\mbox{\tiny Vac.}}^{(1)}(a)=\frac{-\lambda L^{d-1}\pi^2\Omega_{d}^2}{32(2\pi)^{2d}a^{2d-3}}\sum_{n,n'=1}^{\infty}
                                         \big(1+\frac{1}{2}\delta_{n,n'}\big)\omega_{a,n}^{d-2}\omega_{a,n'}^{d-2}, &\hspace{2cm} d=1,3,5,7,9,....
\end{eqnarray}
Based on eqs.\eqref{computing.the.integral.A1.EVEN} and \eqref{computing.the.integral.A1.ODD}, two different types of expression are available for the vacuum energy in even and odd spatial dimensions. Therefore, henceforth, we split the calculation of the Casimir energy into two parts and conduct it in the following separated subsections.

\subsection{Even Dimensions}
As defined in eq.\,\eqref{BSS.Def.}, in the BSS, the vacuum energies of the two configurations\,($\mathcal{A}$ and $\mathcal{B}$ in fig.\,(\ref{BSS.Fig})) should be subtracted from each other. Thus, we have:
\begin{eqnarray}\label{subtraction.BSS.}
     \Delta E_{\mbox{\tiny Vac.}}^{(1)}= E^{(1)}_\mathcal{A}-E^{(1)}_\mathcal{B}=E_{\mbox{\tiny Vac.}}^{(1)}(a)+2E_{\mbox{\tiny Vac.}}^{(1)}(\frac{L-a}{2})-E_{\mbox{\tiny
     Vac.}}^{(1)}(b)-2E_{\mbox{\tiny Vac.}}^{(1)}(\frac{L-b}{2}).
\end{eqnarray}
After substituting eq.\,\eqref{computing.the.integral.A1.EVEN} in eq.\,\eqref{subtraction.BSS.}, we obtain:
\begin{eqnarray}\label{computing.the.integral.EVEN}
   \Delta E_{\mbox{\tiny Vac.}}^{(1)}=\frac{-\lambda L^{d-1}\Omega_{d}^2}{8(2\pi)^{2d}}\sum_{n,n'=1}^{\infty}
                                         \Bigg[\underbrace{\frac{\omega_{a,n}^{d-2}\omega_{n'}^{d-2}}{a^{2d-3}}\ln\omega_{a,n}\ln\omega_{a,n'}}_{\mathcal{U}_{nn'}(a)}
                                         +2\mathcal{U}_{nn'}\big(\frac{L-a}{2}\big)-\{a\to b\}\Bigg]\big(1+\frac{1}{2}\delta_{n,n'}\big), &
                                         d=2,4,6,8,....\nonumber\\
\end{eqnarray}
For all even values of dimension $d$, all summations in eq.\,\eqref{computing.the.integral.EVEN} are divergent. These summations should be regularized and their infinities should be removed. For this purpose and in order to convert the summation forms into an integral form, we used the following form of Abel-Plana Summation Formula\,(APSF):
\begin{eqnarray}\label{APSF}
     \sum_{n=1}^{\infty}\mathcal{F}(n)=\frac{-1}{2}\mathcal{F}(0)+\int_{0}^{\infty}\mathcal{F}(x)dx
     +i\int_{0}^{\infty}\frac{\mathcal{F}(it)-\mathcal{F}(-it)}{e^{2\pi t}-1}dt,
\end{eqnarray}
where the first, second, and last terms on the right-hand side are usually known as \emph{zero}, \emph{integral} and \emph{Branch-cut} term\,(to see more details about the APSF see ref.\,\cite{APSF}). Following the application of the APSF to eq.\,\eqref{computing.the.integral.EVEN}, it is converted to:
\begin{eqnarray}\label{after.APSF.}
     \Delta E_{\mbox{\tiny Vac.}}^{(1)}&=&\frac{-\lambda L^{d-1}\Omega_{d}^2}{8(2\pi)^{2d}a^{2d-3}}\Bigg\{\Bigg[\frac{-1}{2}(ma)^{d-2}\ln
     (ma)+\underbrace{\int_{0}^{\infty}(x^2\pi^2+m^2a^2)^{\frac{d-2}{2}}\ln(x^2\pi^2+m^2a^2)^{\frac{1}{2}}dx}_{\mathcal{I}_1(a)}
     \nonumber \\&+&B_1(a)\Bigg]^2
     +\frac{1}{2}\Bigg[\frac{-1}{2}(m^2a^2)^{d-2}\ln^2(ma)+\underbrace{\int_{0}^{\infty}
     (x^2\pi^2+m^2a^2)^{d-2}\ln^2(x^2\pi^2+m^2a^2)^{\frac{1}{2}}dx}_{\mathcal{I}_2(a)}
     \nonumber\hspace{2.3cm}\\&+&B_2(a)\Bigg]\Bigg\}
     +2\times\left\{a\rightarrow\frac{L-a}{2}\right\}-\big\{a\rightarrow b\big\}-2\times\left\{a\rightarrow\frac{L-b}{2}\right\},
\end{eqnarray}
where $B_1(\alpha)$ and $B_2(\alpha)$ are the Branch-cut terms of APSF. As will be shown, the values of the Branch-cut terms are finite. The terms $\mathcal{I}_1(\alpha)$ and $\mathcal{I}_2(\alpha)$ are the integral terms of APSF, and their values are divergent. All infinities due to these terms should be removed. For this purpose, the first bracket of eq.\,\eqref{after.APSF.} was expanded as:
\begin{eqnarray}\label{after.Apsf.2}
   \Delta E_{\mbox{\tiny Vac.}}^{(1)}&=&\frac{-\lambda L^{d-1}\Omega_{d}^2}{8(2\pi)^{2d}a^{2d-3}}\Bigg\{\Bigg[\frac{1}{4}(ma)^{2d-4}\ln^2(ma)
   +\mathcal{I}^2_1(a)+B_1^2(a) \hspace{4cm}\nonumber\\&+&2\mathcal{I}_1(a)B_1(a)-(ma)^{d-2}\ln(ma)\mathcal{I}_1(a)-(ma)^{d-2}\ln(ma)
   B_1(a)\Bigg]
    +\frac{1}{2}\Bigg[\frac{-1}{2}(m^2a^2)^{d-2}\ln^2(ma)\nonumber\\&+&\mathcal{I}_2(a)+B_2(a)\Bigg]\Bigg\}
     +2\times\left\{a\rightarrow\frac{L-a}{2}\right\}-\big\{a\rightarrow b\big\}-2\times\left\{a\rightarrow\frac{L-b}{2}\right\}.
\end{eqnarray}
Afterwards, to manifest the divergent part of integral $\mathcal{I}_1$, we used the cutoff regularization technique. Therefore, we replaced the upper limit of the integral $\mathcal{I}_1$ by a cutoff value. After computing the integral and expanding the result in the infinite limit of cutoff, we obtain:
\begin{eqnarray}\label{I1(a)}
      \mathcal{I}_1(a)&=&\int_{0}^{\infty}(x^2\pi^2+m^2a^2)^{\frac{d-2}{2}}\ln(x^2\pi^2+m^2a^2)^{\frac{1}{2}}dx\nonumber\\
      &=&\frac{(ma)^{d-1}}{\pi}\Bigg[\ln
      ma\int_{0}^{\Lambda}(\xi^2+1)^{\frac{d-2}{2}}d\xi+\int_{0}^{\Lambda}(\xi^2+1)^{\frac{d-2}{2}}\ln(\xi^2+1)^{\frac{1}{2}}d\xi\Bigg]\nonumber\\
      &&\buildrel \Lambda\to\infty \over{\longrightarrow}\frac{(ma)^{d-1}}{\pi}\Bigg[\ln(ma)
      \Big[\Lambda+\mathcal{O}(\Lambda^3)\Big]+\Big[\frac{\pi}{2}\frac{(d-2)!!}{(d-1)!!}
     +\mathcal{O}(\Lambda)\Big]\Bigg],
\end{eqnarray}
where $\xi=\frac{x\pi}{ma}$. The expansion form displayed in eq.\,\eqref{I1(a)} manifests the divergent parts of integral $\mathcal{I}_1$. Now, we can substitute this expansion form of integral $\mathcal{I}_1$ in the appropriate places in eq.\,\eqref{after.Apsf.2}. The upper limit of integral $\mathcal{I}_1$ in each term of eq.\,\eqref{after.Apsf.2} can be determined differently. We maintain that sufficient degrees of freedom for such type of determination is mathematically available. Therefore, we put the cutoffs $\Lambda_{\mbox{\tiny1(A1)}}$, $\Lambda_{\mbox{\tiny1(A2)}}$, $\Lambda_{\mbox{\tiny1(B1)}}$, and $\Lambda_{\mbox{\tiny1(B2)}}$, on the upper limit of integral $\mathcal{I}_1(\alpha)$ related to regions $A1$, $A2$, $B1$, and $B2$, respectively. Given this setup for cutoffs, the term $2B_1(\alpha)\mathcal{I}_1(\alpha)$ from eq.\,\eqref{after.Apsf.2} is obtained as follows:
\begin{eqnarray}\label{I1(a)--2}
     &&\hspace{-2cm}\frac{1}{a^{2d-3}}\Big[2B_1(a)\mathcal{I}_1(a)\Big]+\frac{2}{\big(\frac{L-a}{2}\big)^{2d-3}}\Big[2B_1\big(\frac{L-a}{2}\big)
     \mathcal{I}_1\big(\frac{L-a}{2}\big)\Big]-\{a\to b\}\nonumber\\
     &=&\frac{2B_1(a)m^{d-1}}{\pi a^{d-2}}\Bigg[\ln(ma) \Big[\Lambda_{\mbox{\tiny1(A1)}}+\mathcal{O}(\Lambda_{\mbox{\tiny1(A1)}}^3)\Big]+\Big[\frac{\pi}{2}\frac{(d-2)!!}{(d-1)!!}
     +\mathcal{O}(\Lambda_{\mbox{\tiny1(A1)}})\Big]\Bigg]\nonumber\\
     &+&\frac{4B_1(\frac{L-a}{2})m^{d-1}}{\pi\big(\frac{L-a}{2}\big)^{d-2}}\Bigg[\ln\big(\frac{m(L-a)}{2}\big)
     \Big[\Lambda_{\mbox{\tiny1(A2)}}+\mathcal{O}(\Lambda_{\mbox{\tiny1(A2)}}^3)\Big]+\Big[\frac{\pi}{2}\frac{(d-2)!!}{(d-1)!!}
     +\mathcal{O}(\Lambda_{\mbox{\tiny1(A2)}})\Big]\Bigg]-\{a\to b\}.
\end{eqnarray}
The number of terms in eq.\,\eqref{I1(a)--2}, which are a function of cutoffs, is divergent when the cutoffs tend to infinity. By determining a relation between cutoffs $\Lambda_{\mbox{\tiny 1(A1)}}$, $\Lambda_{\mbox{\tiny 1(A2)}}$, $\Lambda_{\mbox{\tiny 1(B1)}}$, and $\Lambda_{\mbox{\tiny 1(B2)}}$ in each even dimension, these infinities can be removed from eq.\,\eqref{I1(a)--2}. For instance, in the case of $d=2$, if we determine the cutoffs from relations
$\frac{\Lambda_{\mbox{\tiny 1(B1)}}}{\Lambda_{\mbox{\tiny 1(A1)}}}=\frac{aB_1(a)}{bB_1(b)}\frac{\ln\Lambda_{\mbox{\tiny 1(A1)}}+\ln ma-1}{\ln\Lambda_{\mbox{\tiny 1(B1)}}+\ln mb-1}$ and
$\frac{\Lambda_{\mbox{\tiny 1(B2)}}}{\Lambda_{\mbox{\tiny 1(A2)}}}=\frac{(L-a)B_1(\frac{L-a}{2})}{(L-b)B_1(\frac{L-b}{2})}\frac{\ln\Lambda_{\mbox{\tiny 1(A2)}}+\ln(m(L-a)/2)-1}{\ln\Lambda_{\mbox{\tiny 1(B2)}}+\ln(m(L-b)/2)-1}$, all infinite terms that are functions of cutoffs in eq.\,\eqref{I1(a)--2} will be removed. Finding similar relations between the cutoffs for any other even value of dimensions eliminates all infinities due to the cutoffs. We maintain that in each dimension $d$, enough degrees of freedom is available for this adjustment. Therefore, the remaining finite contribution from eq.\,\eqref{I1(a)--2} becomes:
\begin{eqnarray}\label{remain.2B1I1}
     &&\hspace{-3cm}\frac{1}{a^{2d-3}}\Big[2B_1(a)\mathcal{I}_1(a)\Big]+\frac{2}{\big(\frac{L-a}{2}\big)^{2d-3}}
     \Big[2B_1\big(\frac{L-a}{2}\big)\mathcal{I}_1\big(\frac{L-a}{2}\big)\big]
     -\{a\to b\}\nonumber\\
     &\rightsquigarrow&\frac{B_1(a)m^{d-1}}{a^{d-2}}\frac{(d-2)!!}{(d-1)!!}
     +\frac{2B_1(\frac{L-a}{2})m^{d-1}}{\big(\frac{L-a}{2}\big)^{d-2}}\frac{(d-2)!!}{(d-1)!!}-\{a\to b\}.
\end{eqnarray}
Regarding the term $(m\alpha)^{d-2}\ln(m\alpha)\mathcal{I}_1(\alpha)$ in eq.\,\eqref{after.Apsf.2} after substituting the expansion form of $\mathcal{I}_1$ from
eq.\,\eqref{I1(a)}, we obtain:
\begin{eqnarray}\label{f(0)I1(a)}
     &&\hspace{-1cm}\frac{(ma)^{d-2}\ln(ma)}{a^{2d-3}}\mathcal{I}_1(a)
     +\frac{2\big(\frac{m(L-a)}{2}\big)^{d-2}\ln\big(\frac{m(L-a)}{2}\big)}{\big(\frac{L-a}{2}\big)^{2d-3}}\mathcal{I}_1\big(\frac{L-a}{2}\big)
     -\{a\to b\}\nonumber\\
     &&=\frac{m^{2d-3}\ln ma}{\pi}\Bigg[\ln(ma) \Big[\Lambda_{\mbox{\tiny 2(A1)}}+\mathcal{O}(\Lambda_{\mbox{\tiny 2(A1)}}^3)\Big]+\Big[\frac{\pi}{2}\frac{(d-2)!!}{(d-1)!!}
     +\mathcal{O}(\Lambda_{\mbox{\tiny 2(A1)}})\Big]\Bigg]\nonumber\\
     &&+\frac{2m^{2d-3}\ln\big(\frac{m(L-a)}{2}\big)}{\pi}
     \Bigg[\ln\big(\frac{m(L-a)}{2}\big) \Big[\Lambda_{\mbox{\tiny 2(A2)}}+\mathcal{O}(\Lambda_{\mbox{\tiny 2(A2)}}^3)\Big]+\Big[\frac{\pi}{2}\frac{(d-2)!!}{(d-1)!!}
     +\mathcal{O}(\Lambda_{\mbox{\tiny 2(A2)}})\Big]\Bigg]
     -\{a\to b\}.
\end{eqnarray}
where the cutoffs $\Lambda_{\mbox{\tiny 2(A1)}}$, $\Lambda_{\mbox{\tiny 2(A2)}}$, $\Lambda_{\mbox{\tiny 2(B1)}}$, and $\Lambda_{\mbox{\tiny 2(B2)}}$ are replaced in each region $A1$, $A2$, $B1$, and $B2$, respectively. In the infinite limit of cutoffs, the number of terms in the above expansion is divergent. To remove these infinite terms, an appropriate adjustment is once again required for the value of the cutoffs. This adjustment should be conducted for each dimension, separately. Concerning our example case\,(in dimension $d=2$), by adjusting the cutoffs as a relation $\frac{\Lambda_{\mbox{\tiny2(B1)}}}{\Lambda_{\mbox{\tiny2(A1)}}}=\frac{\ln ma}{\ln mb}\frac{\ln\Lambda_{\mbox{\tiny2(A1)}}+\ln ma-1}{\ln\Lambda_{\mbox{\tiny2(B1)}}+\ln mb-1}$ and
$\frac{\Lambda_{\mbox{\tiny2(B2)}}}{\Lambda_{\mbox{\tiny2(A2)}}}=\frac{\ln(m(L-a)/2)}{\ln(m(L-b)/2)}\frac{\ln\Lambda_{\mbox{\tiny2(A2)}}+\ln(m(L-a)/2)-1}
{\ln\Lambda_{\mbox{\tiny2(B2)}}+\ln(m(L-b)/2)-1}$ for eq.\,\eqref{f(0)I1(a)}, all divergences associated with the cutoffs will be removed. Similarly, by adjusting the cutoffs in the other even dimensions, the remaining finite contribution of eq.\,\eqref{f(0)I1(a)} becomes:
\begin{eqnarray}\label{f(0)I1--remained}
     &&\hspace{-3cm}\frac{(ma)^{d-2}\ln ma}{a^{2d-3}}\mathcal{I}_1(a)
     +\frac{2\big(\frac{m(L-a)}{2}\big)^{d-2}\ln\big(\frac{m(L-a)}{2}\big)}{\big(\frac{L-a}{2}\big)^{2d-3}}\mathcal{I}_1\big(\frac{L-a}{2}\big)
     -\{a\to b\}\nonumber\\
     &&\rightsquigarrow \frac{m^{2d-3}}{2}\frac{(d-2)!!}{(d-1)!!}\Big[\ln ma+2\ln\big(\frac{m(L-a)}{2}\big) -\{a\to b\}\Big].
\end{eqnarray}
Performing a similar scenario for $\mathcal{I}_1^2(\alpha)$ in eq.\,\eqref{after.Apsf.2} eliminates all contributions of this term from eq.\,\eqref{after.Apsf.2}. In fact, the term $\mathcal{I}_1^2(\alpha)$ will not leave any contribution in the Casimir energy expression. The term $\mathcal{I}_2(\alpha)$ in eq.\,\eqref{after.Apsf.2} is still divergent. To regularize this term and remove its infinity from eq.\,\eqref{after.Apsf.2}, we once again employed the cutoff regularization technique. So, we have:
\begin{eqnarray}\label{I2(a)}
      \frac{1}{a^{2d-3}}\mathcal{I}_2(a)&=&\frac{1}{a^{2d-3}}\int_{0}^{\infty}(x^2\pi^2+m^2a^2)^{d-2}\ln^2(x^2\pi^2+m^2a^2)^{\frac{1}{2}}dx
      =\frac{m^{2d-3}}{\pi}\ln(ma)\int_{0}^{\Lambda}(\xi^2+1)^{d-2}d\xi\nonumber\\
      &+&\frac{m^{2d-3}}{\pi}\ln(ma)\int_{0}^{\Lambda}(\xi^2+1)^{d-2}\ln(\xi^2+1)d\xi
      +\frac{m^{2d-3}}{4\pi}\int_{0}^{\infty}(\xi^2+1)^{d-2}\ln^2(\xi^2+1)d\xi,
\end{eqnarray}
where $\xi=\frac{x\pi}{ma}$. Note that the last term automatically cancels out when the vacuum energies written in eq.\,\eqref{after.Apsf.2} are subtracted; therefore, we do not need to change the upper limit of this term on the right-hand side of the above equation. Regarding the first two terms on the right-hand side of eq.\,\eqref{I2(a)}, we should calculate the integrals and expand the results in the infinite limit of cutoffs. Now, we substitute this expansion form in eq.\,\eqref{after.Apsf.2} and for the term $\frac{1}{2}\mathcal{I}_2(\alpha)$ we obtain:
\begin{eqnarray}\label{I2(a)---2}
       &&\frac{1}{a^{2d-3}}\mathcal{I}_2(a)+\frac{2}{\big(\frac{L-a}{2}\big)^{2d-3}}\mathcal{I}_2\big(\frac{L-a}{2}\big)-\{a\to b\}\nonumber\\
       &&=\frac{m^{2d-3}}{\pi}\ln(ma)\Bigg[\Lambda_{\mbox{\tiny3(A1)}}+\mathcal{O}(\Lambda_{\mbox{\tiny3(A1)}}^{3})\Bigg]
       +\frac{m^{2d-3}}{\pi}\ln(ma)\Bigg[\frac{\pi(2d-4)!!}{(2d-3)!!}+\mathcal{O}(\Lambda_{\mbox{\tiny3(A1)}})\Bigg]\nonumber\\
       &&+\frac{m^{2d-3}}{4\pi}\int_{0}^{\infty}(\xi^2+1)^{d-2}\ln^2(\xi^2+1)d\xi
       +\frac{2m^{2d-3}}{\pi}\ln\big(\frac{m(L-a)}{2}\big)\Bigg[\Lambda_{\mbox{\tiny3(A2)}}+\mathcal{O}(\Lambda_{\mbox{\tiny3(A2)}}^{3})\Bigg]\nonumber\\
       &&+\frac{2m^{2d-3}}{\pi}\ln\big(\frac{m(L-a)}{2}\big)\Bigg[\frac{\pi(2d-4)!!}{(2d-3)!!}
       +\mathcal{O}(\Lambda_{\mbox{\tiny3(A2)}})\Bigg]+\frac{m^{2d-3}}{2\pi}\int_{0}^{\infty}(\xi^2+1)^{d-2}\ln^2(\xi^2+1)d\xi
       -\{a\to b\}\nonumber\\
       &&\rightsquigarrow  \frac{m^{2d-3}(2d-4)!!}{(2d-3)!!}\left[\ln(ma)+2\ln\big(\frac{m(L-a)}{2}\big)-\{a\to b\}\right].
\end{eqnarray}
In this equation, the cutoffs $\Lambda_{\mbox{\tiny3(A1)}}$, $\Lambda_{\mbox{\tiny3(A2)}}$, $\Lambda_{\mbox{\tiny3(B1)}}$, and $\Lambda_{\mbox{\tiny3(B2)}}$ are considered for terms pertaining to regions $A1$, $A2$, $B1$, and $B2$, respectively. In each even spatial dimension, the proper determination of cutoffs eliminates all infinite terms as functions of the cutoffs in eq.\,\eqref{after.Apsf.2}$^1$.\footnotetext[1]{For example, in $d=2$ by determining of the cutoffs from the following relations all divergent parts due to the $\mathcal{I}_2$ would be eliminated from eq.\,\eqref{after.Apsf.2}:
\begin{eqnarray}\label{cut.adjust.I2}
      \frac{\Lambda_{\mbox{\tiny3(B1)}}}{\Lambda_{\mbox{\tiny3(A1)}}}=\frac{\ln(ma)\big[2\ln\Lambda_{\mbox{\tiny3(A1)}}-1\big]}{\ln(mb)\big[2\ln\Lambda_{\mbox{\tiny3(B1)}}-1\big]},\hspace{2.5cm}
      \frac{\Lambda_{\mbox{\tiny3(B2)}}}{\Lambda_{\mbox{\tiny3(A2)}}}=\frac{\ln(m(L-a)/2)\big[2\ln\Lambda_{\mbox{\tiny3(A2)}}-1\big]}{\ln(m(L-b)/2)\big[2\ln\Lambda_{\mbox{\tiny3(B2)}}-1\big]}.\nonumber
\end{eqnarray}}
\par
\begin{figure}[th] \hspace{-1cm}\includegraphics[width=8.5cm]{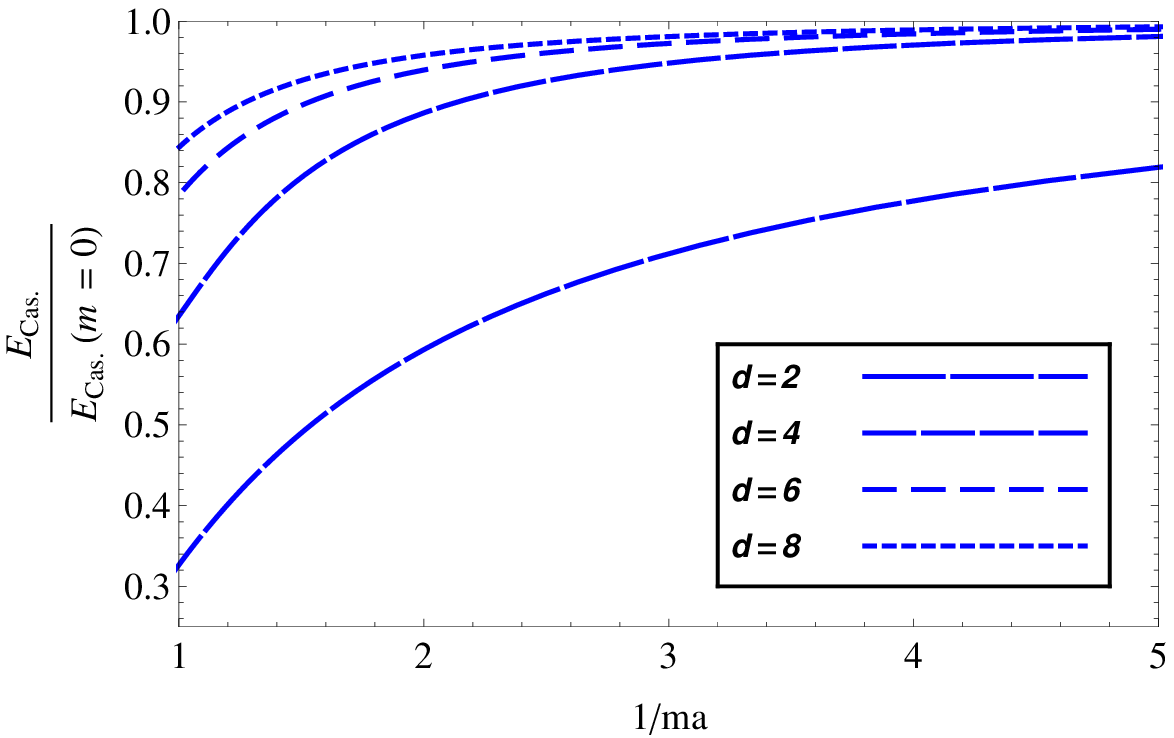}\hspace{0.5cm}\includegraphics[width=8.5cm]{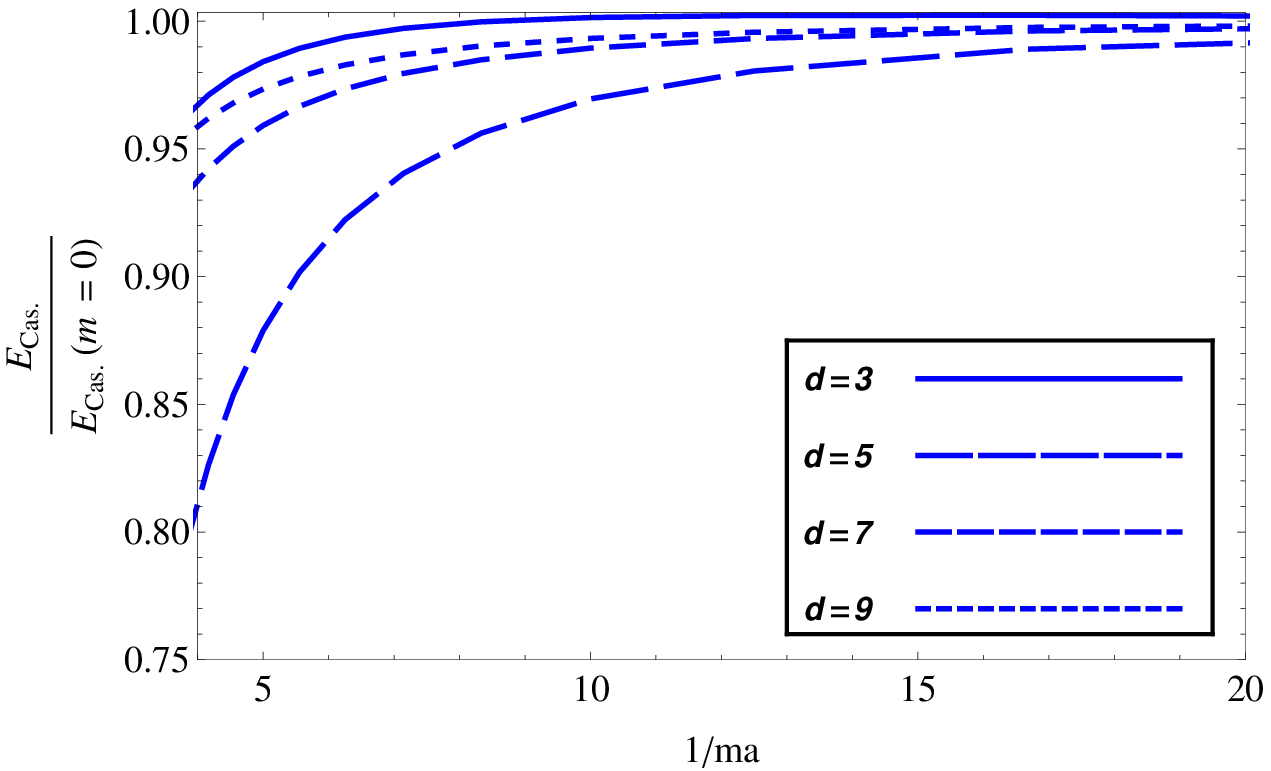}
\caption{\label{fig.ratio.Odd&Even} \small
  The ratio of the first-order radiative correction to the Casimir energy density for massive scalar field compared to the massless one as a function of $1/ma$. The Left(Right) figure presents the sequence of plots for even(odd) dimensions. The coupling constant value of all plots is $\lambda=0.1$.}
\end{figure}

As mentioned earlier, the value of Branch-cut terms $B_1(\alpha)$ and $B_2(\alpha)$ in eq.\,\eqref{after.Apsf.2} is finite. To obtain an expression for the Branch-cut term $B_1(a)$ we have:
\begin{eqnarray}\label{B1(a)-EvenDim}
       B_1(a)&=&i\int_{0}^{\infty}
       \frac{\Big((it\pi)^2+m^2a^2\Big)^{\frac{d-2}{2}}\ln\Big((it\pi)^2+m^2a^2\Big)^{\frac{1}{2}}
       -\Big((-it\pi)^2+m^2a^2\Big)^{\frac{d-2}{2}}\ln\Big((-it\pi)^2+m^2a^2\Big)^{\frac{1}{2}}}
       {e^{2\pi t}-1}dt\nonumber\\
       &=&(ma)^{d-1}(-1)^{\frac{d}{2}}\int_{1}^{\infty}\frac{(\eta^2-1)^{\frac{d-2}{2}}}{e^{2ma\eta}-1}d\eta,
\end{eqnarray}
where $\eta=\frac{\pi t}{ma}$. After expanding the denominator of the integrand, we obtain the final expression of $B_1(a)$ as:
\begin{eqnarray}\label{B1(a)--Final form--Even Dim}
       B_1(a)=\frac{(-1)^{\frac{d}{2}}(ma)^{d-1}}{\sqrt{\pi}}\Gamma\left(d/2\right)\sum_{j=1}^{\infty}\frac{K_{\frac{d-1}{2}}(2maj)}{(maj)^{\frac{d-1}{2}}},
\end{eqnarray}
where $K_\nu(\alpha)$ is the modified Bessel function. For the Branch-cut term $B_2(a)$, we have:
\begin{eqnarray}\label{B2(a)-EvenDim}
        B_2(a)&=&i\int_{0}^{\infty}
       \frac{((it\pi)^2+m^2a^2)^{d-2}\ln^2((it\pi)^2+m^2a^2)^{\frac{1}{2}}-((-it\pi)^2+m^2a^2)^{d-2}\ln^2((-it\pi)^2+m^2a^2)^{\frac{1}{2}}}
       {e^{2\pi t}-1}dt\nonumber\\&=&-(ma)^{2d-3}\ln(m^2a^2)\int_{1}^{\infty}\frac{(\eta^2-1)^{d-2}}{e^{2ma\eta}-1}d\eta
       -(ma)^{2d-3}\int_{1}^{\infty}\frac{(\eta^2-1)^{d-2}\ln(\eta^2-1)}{e^{2ma\eta}-1}d\eta,
\end{eqnarray}
where $\eta=\frac{\pi t}{ma}$. After expanding the denominator of integrands and computing all integrations, the final answer for $B_2(a)$ becomes:
\begin{eqnarray}\label{B2(a)---FinalAns--EvenDim}
       B_2(a)&=&\frac{-2\ln(ma)(ma)^{2d-3}\Gamma(d-1)}{\sqrt{\pi}}\sum_{j=1}^{\infty}\frac{K_{d-3/2}(2maj)}{(maj)^{d-3/2}}\nonumber\\
       &+&\frac{\Gamma(d-1)(ma)^{2d-3}}{\sqrt{\pi}} \sum_{j=1}^{\infty}\frac{K_{d-3/2}(2maj)\big[\ln(maj)-\mbox{PolyLog}(0,d-1)\big]-\partial_\nu
       K_\nu(2maj)\Big|_{\nu=d-3/2}}{(maj)^{d-3/2}},
\end{eqnarray}
where $K_\nu(\alpha)$ is the modified Bessel function and,
\begin{eqnarray}\label{polylog}
  \mbox{PolyLog}(0,z)=\int_{0}^{\infty}\left(\frac{e^{-t}}{t}-\frac{1}{t(1+t)^z}\right)dt.
\end{eqnarray}
Through the use of eqs.\,\eqref{remain.2B1I1}, \eqref{f(0)I1--remained}, and \eqref{I2(a)---2}, the expression $\Delta E^{(1)}_{\mbox{\tiny Vac.}}$ given in eq.\,\eqref{after.Apsf.2} is converted to:
\begin{eqnarray}\label{Before limit.Final.Cas.Energy.EvenDim}
  \Delta E^{(1)}_{\mbox{\tiny Vac.}}&=&\frac{-\lambda L^{d-1}\Omega_d^2}{8(2\pi)^d}\Big[\mathcal{E}(a)+2\mathcal{E}\big(\frac{L-a}{2}\big)-\{a\to b\}\Big],
\end{eqnarray}
where
\begin{eqnarray}\label{Before limit.Final.Cas.Energy.EvenDim.2}
      \mathcal{E}(\alpha)&=&\frac{B_1^2(\alpha)}{\alpha^{2d-3}}
      +\frac{m^{d-1}(d-2)!!}{\alpha^{d-2}(d-1)!!}B_1(\alpha)-\frac{m^{2d-3}\ln(m\alpha)}{2}\frac{(d-2)!!}{(d-1)!!}
      \nonumber\\&&\hspace{1.5cm}-\frac{m^{d-2}\ln(m\alpha)B_1(\alpha)}{\alpha^{d-1}}
      +\frac{m^{2d-3}\ln(m\alpha)}{2}\frac{(2d-4)!!}{(2d-3)!!}+\frac{B_2(\alpha)}{2\alpha^{2d-3}}.
\end{eqnarray}
In the final step, using eq.\,\eqref{BSS.Def.}, the limits $L/b\to\infty$ and $b/a\to\infty$ should be applied. After applying these limits to eq.\,\eqref{Before limit.Final.Cas.Energy.EvenDim} for any values of mass $m\neq0$, all contributions associated with regions $A2$, $B1$, and $B2$ are eliminated. Therefore, the final expression of the total Casimir energy regarding the massive scalar field confined between two parallel plates with distance $a$ in even spatial dimensions becomes:
\begin{eqnarray}\label{Final.Cas.Energy.EvenDim}
      E^{(1)}_{\mbox{\tiny Cas.,Even}}(a)&=&\frac{-\lambda L^{d-1}\Omega_d^2}{8(2\pi)^d}\Bigg\{\frac{B_1^2(a)}{a^{2d-3}}
      +\frac{m^{d-1}(d-2)!!}{a^{d-2}(d-1)!!}B_1(a)-\frac{m^{2d-3}\ln(ma)}{2}\frac{(d-2)!!}{(d-1)!!}
      \nonumber\\&&\hspace{3.5cm}-\frac{m^{d-2}\ln(ma)B_1(a)}{a^{d-1}}
      +\frac{m^{2d-3}\ln(ma)}{2}\frac{(2d-4)!!}{(2d-3)!!}+\frac{B_2(a)}{2a^{2d-3}}\Bigg\}.
\end{eqnarray}
The first-order radiative correction to the Casimir energy between a pair of plates with Dirichlet boundary condition in every even spatial dimension was reported in\,\cite{cavalcanti.1,cavalcanti.2,cavalcanti.3}. While their reported result was infinite, our obtained result, written in eq.\,\eqref{Final.Cas.Energy.EvenDim}, was convergent concerning all even spatial dimensions. The main source of this difference can be attributed to the type of the renormalization program implemented in the problem. In the present study, the counterterms used in the renormalization program were position-dependent and consistent with the imposed boundary condition. However, the counterterms used in the previous works were free counterterms. Our definition of free counterterm is the one where in Minkowski space\,(free space) is used. In order to verify the consistency of the obtained result in eq.\,\eqref{Final.Cas.Energy.EvenDim}, we resorted to the first-order computation of the Casimir energy for the massless scalar field. The BSS was not used in calculating the Casimir energy for the massless field; the calculations were only performed using the analytic continuation technique so as to create more confidence in checking the consistency of results between the massive and massless cases. Therefore, to obtain the first-order radiative correction to the Casimir energy for the massless scalar field we go back to eq.\,\eqref{computing.the.integral.A1.EVEN} and set the mass parameter $m=0$. Therefore, we obtain:
\begin{eqnarray}\label{computing.the.integral.A1.EVEN.MasslessCase}
E_{\mbox{\tiny Vac.}}^{(1)}(a)=\frac{-\lambda L^{d-1}\Omega_{d}^2}{8(2\pi)^{2d}a^{2d-3}}\sum_{n,n'=1}^{\infty}
                                        \big(1+\frac{1}{2}\delta_{n,n'}\big)(n\pi)^{d-2}(n'\pi)^{d-2}\ln(n\pi)\ln(n'\pi), &\hspace{1.5cm}  d=2,4,6,8,....
\end{eqnarray}
Utilizing the analytic continuation technique in calculating the above summation, the radiative correction to the Casimir energy for the massless scalar field between two parallel plates in $d$ spatial dimensions becomes:
\begin{eqnarray}\label{Final.Cas.Energy.EvenDim.Massless}
E^{(1)}_{\mbox{\tiny Cas.,Even}}(m=0,a)=&&\frac{-\lambda L^{d-1}\Omega_d^2}{8(2\pi)^d a^{2d-3}}
\Bigg[\bigg(\zeta(2-d)\ln\pi-\zeta'(2-d)\bigg)^2\nonumber\\
&&\hspace{2cm}+\frac{1}{2}\bigg(\zeta(2-4d)\ln^2\pi-2\zeta'(4-2d)\ln\pi+\zeta''(4-2d)\bigg)
\Bigg],
\end{eqnarray}
where $\zeta(\alpha)$ is the Riemann zeta function. Radiative correction to the Casimir energy for massive and massless scalar fields between two parallel
plates with Dirichlet boundary condition in two spatial dimensions by position-dependent counterterms was reported in\,\cite{2D-Man}. Our results of massive and massless scalar fields\,(eqs.\,\eqref{Final.Cas.Energy.EvenDim} and \eqref{Final.Cas.Energy.EvenDim.Massless}) in the special case\,($d=2$) are exactly in line with the foregoing work. Fig.\,(\ref{fig.ratio.Odd&Even}) plots the ratio of the Casimir energy correction values in two massive and massless cases as a function of $1/ma$ for $d=\{2,4,6,8\}$. This figure satisfies the physical expectations and shows that there is a good consistency between the results obtained from eqs.\,\eqref{Final.Cas.Energy.EvenDim} and \eqref{Final.Cas.Energy.EvenDim.Massless}. To obtain the radiative correction to the Casimir energy for the Lorentz-violating scalar field\,($\beta\neq0$), the calculation procedure is similar to the case of $\beta=0$. The Green's function expression associated with each violating direction of the Lorentz symmetry breaking given in eqs.\,\eqref{Green.function.TLDBC}, \eqref{Green.function.SLPADBC}, and \eqref{Green.function.SLPRDBC} should be substituted with the vacuum energy expression given in eq.\,\eqref{VacuumEn.firstorder.}. The calculation procedure including the BSS and cutoff regularization technique is similar to the case of $\beta=0$; therefore, the BSS was once again employed as a regularization technique. Finally, the radiative correction to the Casimir energy for the massive scalar field for each violated direction of the Lorentz symmetry is obtained as:
\begin{eqnarray}\label{Final.Cas.Energy.EvenDim.LB.TL&SLPA.Massive}
     E^{(1)}_{\mbox{\tiny Cas.,TL}}(a)&=&\frac{1}{1+\beta}E^{(1)}_{\mbox{\tiny Cas.,Even}}(a),\nonumber\\
     E^{(1)}_{\mbox{\tiny Cas.,SL-Par}}(a)&=&\frac{1}{1-\beta}E^{(1)}_{\mbox{\tiny Cas.,Even}}(a),\nonumber\\
     E^{(1)}_{\mbox{\tiny Cas.,SL-Perp}}(a)&=&E^{(1)}_{\mbox{\tiny Cas.,Even}}(\tilde{a}).
\end{eqnarray}
where $\tilde{a}=\frac{a}{\sqrt{1-\beta}}$. To obtain the radiative correction to the Casimir energy for the massless scalar field in Lorentz-violating system, we return to eqs.\,\eqref{Green.function.TLDBC}, \eqref{Green.function.SLPADBC}, and \eqref{Green.function.SLPRDBC} and set the mass parameter $m=0$. Afterwards, for each violated direction of the Lorentz symmetry breaking, the associated Green's function expression is substituted in eq.\,\eqref{VacuumEn.firstorder.}; by applying the analytic continuation technique to the vacuum energy, we have:
\begin{eqnarray}\label{Final.Cas.Energy.EvenDim.LB.TL&SLPA.Massless}
     E^{(1)}_{\mbox{\tiny Cas.,TL}}(m=0,a)&=&\frac{1}{1+\beta}E^{(1)}_{\mbox{\tiny Cas.,Even}}(m=0,a),\nonumber\\
     E^{(1)}_{\mbox{\tiny Cas.,SL-Par}}(m=0,a)&=&\frac{1}{1-\beta}E^{(1)}_{\mbox{\tiny Cas.,Even}}(m=0,a),\nonumber\\
     E^{(1)}_{\mbox{\tiny Cas.,SL-Perp}}(m=0,a)&=&E^{(1)}_{\mbox{\tiny Cas.,Even}}(m=0,\tilde{a}).
\end{eqnarray}
Figs.\,(\ref{fig.LB.Odd&Even.TL&SLPA&SLPER}a), (\ref{fig.LB.Odd&Even.TL&SLPA&SLPER}b), and (\ref{fig.LB.Odd&Even.TL&SLPA&SLPER}c) plot the radiative correction to the Casimir energy as a function of the distance of plates for $d=\{2,4,6,8\}$ and $\beta=\{0,0.1,0.2,0.5\}$. The sequence of plots for three violated directions of the Lorentz symmetry breaking\,(TL, SL-Par and SL-Perp) is displayed in separate figures. Fig.\,(\ref{fig.LB.Odd&Even.TL&SLPA&SLPER}) shows that the radiative correction to the Casimir energy was positive for all even dimensions. The effect of Lorentz symmetry breaking on the Casimir energy value in the case of SL-Par was higher than the other violated direction of the Lorentz symmetry breaking. Fig.\,(\ref{fig.LB.Odd&Even.TL&SLPA&SLPER}) further shows that Lorentz violating in the time-like case had the minimum effect on the value of the Casimir energy.

\subsection{Odd Dimensions}
To compute the radiative correction to the Casimir energy for massive scalar field between two parallel plates in odd spatial dimensions, we substituted eq.\,\eqref{computing.the.integral.A1.ODD} in eq.\eqref{subtraction.BSS.}. Therefore, we obtain:
\begin{eqnarray}\label{computing.the.integral.ODD}
   \Delta E_{\mbox{\tiny Vac.}}^{(1)}=\frac{-\lambda L^{d-1}\pi^2\Omega_{d}^2}{32(2\pi)^{2d}}\sum_{n,n'=1}^{\infty}
                                         \Bigg[\underbrace{\frac{\omega_{a,n}^{d-2}\omega_{a,n'}^{d-2}}{a^{2d-3}}}_{\mathcal{T}_{nn'}(a)}
                                         +2\mathcal{T}_{nn'}\big(\frac{L-a}{2}\big)-\{a\to b\}\Bigg]\big(1+\frac{1}{2}\delta_{n,n'}\big), & d=3,5,7,9,....
\end{eqnarray}
The first sum is divergent in the vacuum energy given in eq.\,\eqref{computing.the.integral.A1.ODD} regarding any odd spatial dimensions, while the second one converges for dimension $d=1$ and diverges in the other odd spatial dimensions. Such different behavior in the second summation in eq.\,\eqref{computing.the.integral.A1.ODD} made us present a separate calculation for the radiative correction to the Casimir energy in $d=1$. Note that \cite{1D-Reza} previously calculated the radiative correction to the Casimir energy in $1+1$ space-time dimensions using position-dependent counterterms. Therefore, we did not repeat all the calculations here; in the Appendix\,\ref{appendix.1Dim}, by use of their results, we only obtained the radiative correction to the Casimir energy for the Lorentz-violating scalar field in $1+1$ dimensions. In the following, using eq.\,\eqref{computing.the.integral.ODD}, we continue the calculations for other odd spatial dimensions\,($d\neq1$). For this purpose, and in order to regularize the divergences appearing in eq.\,\eqref{computing.the.integral.ODD}, the APSF given in eq.\eqref{APSF} was applied. The APSF converts all summation forms of eq.\,\eqref{computing.the.integral.ODD} into the integral form. Therefore, we obtain:
\begin{eqnarray}\label{after.APSF.oddDim}
     \Delta E_{\mbox{\tiny Vac.}}^{(1)}&=&\frac{-\lambda L^{d-1}\pi^2\Omega_d^2}{32(2\pi)^{2d}a^{2d-3}}
     \Bigg\{\Big[\frac{-1}{2}(ma)^{d-2}+\underbrace{\int_{0}^{\infty}(x^2\pi^2+m^2a^2)^{\frac{d-2}{2}}dx}_{\mathcal{J}_1(a)}
     +\mathcal{B}_1(a)\Big]^2\nonumber \\
     &+&\frac{1}{2}\left[\frac{-1}{2}(m^2a^2)^{d-2}+\underbrace{\int_{0}^{\infty}(x^2\pi^2+m^2a^2)^{d-2}dx}_{\mathcal{J}_2(a)}+\mathcal{B}_2(a)\right]\Bigg\}\nonumber\\
     &+&2\times\left\{a\rightarrow\frac{L-a}{2}\right\}-\{a\rightarrow b\}-2\times\left\{a\rightarrow\frac{L-b}{2}\right\},\nonumber\\
\end{eqnarray}
where $\mathcal{B}_1(\alpha)$ and $\mathcal{B}_2(\alpha)$ are the Branch-cut terms of APSF and their values are finite. After expanding the first bracket in eq.\,\eqref{after.APSF.oddDim}, it is converted into:
\begin{eqnarray}\label{after.Apsf.2.oddDim}
   \Delta E_{\mbox{\tiny Vac.}}^{(1)}&=&\frac{-\lambda L^{d-1}\pi^2\Omega_d^2}{32(2\pi)^{2d}a^{2d-3}}
   \Bigg\{\frac{1}{4}(ma)^{2d-4}+\mathcal{J}^2_1(a)+\mathcal{B}_1^2(a)+2\mathcal{J}_1(a)\mathcal{B}_1(a)-(ma)^{d-2}\mathcal{J}_1(a)
    \nonumber\\ &-&(ma)^{d-2}\mathcal{B}_1(a)-\frac{1}{4}(m^2a^2)^{d-2}+\frac{1}{2}\mathcal{J}_2(a)+\frac{1}{2}\mathcal{B}_2(a)\Bigg\}\nonumber\\
    &+&2\times\left\{a\rightarrow\frac{L-a}{2}\right\}-\{a\rightarrow b\}-2\times\left\{a\rightarrow\frac{L-b}{2}\right\}.
\end{eqnarray}
The integral terms $\mathcal{J}_1(\alpha)$ and $\mathcal{J}_2(\alpha)$ are divergent. The same as the procedure conducted in the case of even dimensions, we used the cutoff regularization technique to remove the divergences originating from these integral terms. Therefore, for the term $2\mathcal{J}_1(\alpha)\mathcal{B}_1(\alpha)$ from eq.\,\eqref{after.Apsf.2.oddDim} we have:
\begin{eqnarray}\label{2J1B1}
       &&\frac{1}{a^{2d-3}}2\mathcal{J}_1(a)\mathcal{B}_1(a)+\frac{1}{\big(\frac{L-a}{2}\big)^{2d-3}}4\mathcal{J}_1\Big(\frac{L-a}{2}\Big)B_1\Big(\frac{L-a}{2}\Big)-\{a\to
       b\}\nonumber\\
       &&=2\mathcal{B}_1(a)\frac{m^{d-1}}{\pi a^{d-2}} \int_{0}^{\Lambda_{\mbox{\tiny4(A1)}}}\left(\xi^2+1\right)^{\frac{d-2}{2}}d\xi
       +4\mathcal{B}_1\Big(\frac{L-a}{2}\Big)\frac{m^{d-1}}{\pi\big(\frac{L-a}{2}\big)^{d-2}} \int_{0}^{\Lambda_{\mbox{\tiny4(A2)}}}\left(\xi^2+1\right)^{\frac{d-2}{2}}d\xi-\{a\to
       b\},
\end{eqnarray}
where $\xi=\frac{x\pi}{m\alpha}$. For each dimension, all integrals in eq.\,\eqref{2J1B1} should be computed and their results expanded in the infinite limit of cutoffs. A proper adjusting for the value of cutoffs helps remove the divergent parts of expansions via subtraction procedure defined by the BSS. As an example for $d=3$, the eq.\,\eqref{2J1B1} is converted to:
\begin{eqnarray}\label{2B1J1-2}
     &&\hspace{-2cm}\frac{1}{a^{2d-3}}2\mathcal{J}_1(a)\mathcal{B}_1(a)+\frac{1}{\big(\frac{L-a}{2}\big)^{2d-3}}4\mathcal{J}_1\Big(\frac{L-a}{2}\Big)
     \mathcal{B}_1\Big(\frac{L-a}{2}\Big)-\{a\to
     b\}\nonumber\\
     &&=2\mathcal{B}_1(a)\frac{m^{d-1}}{\pi a^{d-2}}\Bigg[\frac{1+\ln4}{4}+\frac{\ln\Lambda_{\mbox{\tiny4(A1)}}+\Lambda_{\mbox{\tiny4(A1)}}^2}{2}+\mathcal{O}(\Lambda_{\mbox{\tiny4(A1)}}^{-2})\Bigg]\nonumber\\
     &&+4\mathcal{B}_1\Big(\frac{L-a}{2}\Big)\frac{m^{d-1}}{\pi\big(\frac{L-a}{2}\big)^{d-2}}
     \Bigg[\frac{1+\ln4}{4}+\frac{\ln\Lambda_{\mbox{\tiny4(A2)}}+\Lambda_{\mbox{\tiny4(A2)}}^2}{2}+\mathcal{O}(\Lambda_{\mbox{\tiny4(A2)}}^{-2})\Bigg]-\{a\to b\}.
\end{eqnarray}
Adjusting the cutoffs as the relation $\frac{\ln\Lambda_{\mbox{\tiny4(B1)}}+\Lambda_{\mbox{\tiny4(B1)}}^2}{\ln\Lambda_{\mbox{\tiny4(A1)}}+\Lambda_{\mbox{\tiny4(A1)}}^2}=\frac{b\mathcal{B}_1(a)}{a\mathcal{B}_1(b)}$ and
$\frac{\ln\Lambda_{\mbox{\tiny4(B2)}}+\Lambda_{\mbox{\tiny4(B2)}}^2}{\ln\Lambda_{\mbox{\tiny4(A2)}}+\Lambda_{\mbox{\tiny4(A2)}}^2}=\frac{(L-b)\mathcal{B}_1(\frac{L-a}{2})}{(L-a)\mathcal{B}_1(\frac{L-b}{2})}$ leads to the removal of all divergent terms in eq.\,\eqref{2B1J1-2}. This procedure can further be performed in other odd dimensions\,($d\neq1$). Therefore, the remaining finite terms from eq.\,\eqref{2B1J1-2} for any odd dimension $d$ become:
\begin{eqnarray}\label{2J1B1-2}
       &&\hspace{-2cm}\frac{1}{a^{2d-3}}2\mathcal{J}_1(a)\mathcal{B}_1(a)+\frac{1}{\big(\frac{L-a}{2}\big)^{2d-3}}4\mathcal{J}_1\Big(\frac{L-a}{2}\Big)\mathcal{B}_1\Big(\frac{L-a}{2}\Big)-\{a\to
       b\}\nonumber\\
       &&\rightsquigarrow \Big[\mathcal{H}(d)+\frac{(d-2)!!}{2(d-1)!!}\ln4\Big]
       \Bigg\{2\mathcal{B}_1(a)\frac{m^{d-1}}{\pi a^{d-2}}+4\mathcal{B}_1\Big(\frac{L-a}{2}\Big)\frac{m^{d-1}}{\pi\big(\frac{L-a}{2}\big)^{d-2}}-\{a\to b\}\Bigg\},
\end{eqnarray}
where the values of function $\mathcal{H}(d)$ are listed in Table\,\ref{tableH(d)}.
\begin{table}[th]
   \begin{tabular}{|cc||cc|}
  \hline
 \hspace{0.5cm} $d$    &    \hspace{1cm} $\dd\mathcal{H}(d)$   \hspace{0.5cm}\vspace{0.1cm}&\hspace{0.1cm} $d$ & \hspace{1cm}$\dd\mathcal{H}(d)$
 \hspace{1cm}  \\ \hline
 \hspace{0.5cm} 3    &      \hspace{1cm} $\frac{1}{4}$         \hspace{0.5cm}\vspace{0.1cm}&\hspace{0.1cm} 13 & \hspace{1cm}$\frac{11319}{40960}
 $\hspace{1cm} \\
 \hspace{0.5cm} 5    &      \hspace{1cm} $\frac{9}{32}$        \hspace{0.5cm}\vspace{0.1cm}&\hspace{0.1cm} 15 & \hspace{1cm}$\frac{155727}{573440}$
 \hspace{1cm}              \\
 \hspace{0.5cm} 7    &      \hspace{1cm} $\frac{55}{192}$      \hspace{0.5cm}\vspace{0.1cm}&\hspace{0.1cm} 17 & \hspace{1cm}$\frac{979407}{3670016}$
 \hspace{1cm}    \\
 \hspace{0.5cm} 9    &      \hspace{1cm} $\frac{875}{3072}$    \hspace{0.5cm}\vspace{0.1cm}&\hspace{0.1cm} .. & \hspace{1cm}  $..$ \hspace{1cm}
 \\
 \hspace{0.5cm} 11   &      \hspace{1cm} $\frac{2877}{10240}$    \hspace{0.5cm}\vspace{0.1cm}&\hspace{0.1cm}      & \hspace{1cm} \hspace{1cm}
 \\ \hline
  \end{tabular}\caption{\label{tableH(d)}
        Values of function $\mathcal{H}(d)$ as a function of odd spatial dimensions $d\neq1$.}
\end{table}
\par
For the term $\mathcal{J}^2_1(\alpha)$ in eq.\,\eqref{after.Apsf.2.oddDim}, we have:
\begin{eqnarray}\label{J1(a).oddDim}
     &&\hspace{-2cm} \frac{1}{a^{2d-3}}\mathcal{J}^2_1(a)+2\frac{1}{(\frac{L-a}{2})^{2d-3}}\mathcal{J}^2_1\left(\frac{L-a}{2}\right)-\{a\to b\}\nonumber\\
      &&=\frac{(ma)^{2d-2}}{\pi^2a^{2d-3}}\left(\int_{0}^{\infty}(\xi^2+1)^{\frac{d-2}{2}}d\xi\right)^2
      +2\frac{(m(\frac{L-a}{2}))^{2d-2}}{\pi^2(\frac{L-a}{2})^{2d-3}}\left(\int_{0}^{\infty}(\xi^2+1)^{\frac{d-2}{2}}d\xi\right)^2-\{a\rightarrow
      b\}\nonumber\\
      &&=\left[a+2\frac{L-a}{2}-b-2\frac{L-b}{2}\right]\left(\frac{m^{d-1}}{\pi}\int_{0}^{\infty}(\xi^2+1)^{\frac{d-2}{2}}d\xi\right)^2=0,
\end{eqnarray}
where $\xi=\frac{x \pi}{m\alpha}$. As shown in eq.\,\eqref{J1(a).oddDim}, using the BSS, all expressions of $\mathcal{J}^2_1(\alpha)$ automatically cancel out one another; therefore, no contribution from $\mathcal{J}^2_1(\alpha)$ remains in eq.\,\eqref{after.Apsf.2.oddDim}. Similarly, for the term $(m\alpha)^{d-2}\mathcal{J}_1(\alpha)$ in eq.\,\eqref{after.Apsf.2.oddDim} we obtain:
\begin{eqnarray}\label{mJ1.OddDim}
      &&\hspace{-2cm}\frac{1}{a^{2d-3}}(ma)^{d-2}\mathcal{J}_1(a)+2\frac{1}{(\frac{L-a}{2})^{2d-3}}\Big(\frac{m(L-a)}{2}\Big)^{d-2}\mathcal{J}_1\Big(\frac{L-a}{2}\Big)-\{a\to
      b\}\nonumber\\
      &&=\left[\frac{(ma)^{2d-3}}{\pi a^{2d-3}}+2\frac{\left(m(L-a)\right)^{2d-3}}{\pi(L-a)^{2d-3}}-\{a\to
      b\}\right]\int_{0}^{\infty}(\xi^2+1)^{\frac{d-2}{2}}d\xi=0.
\end{eqnarray}
The integral $\mathcal{J}_2(\alpha)$ is also divergent. However, the remaining contribution of this term following the subtraction process of BSS is exactly zero. Thus, we have:
\begin{eqnarray}\label{J2}
       \frac{1}{a^{2d-3}}\mathcal{J}_2(a)+2\frac{1}{\big(\frac{L-a}{2}\big)^{2d-3}}\mathcal{J}_2\Big(\frac{L-a}{2}\Big)-\{a\to b\}\hspace{6cm}\nonumber\\
        =\left[\frac{(ma)^{2d-3}}{\pi a^{2d-3}}+2\frac{\left(m(L-a)\right)^{2d-3}}{\pi(L-a)^{2d-3}}-\{a\to
        b\}\right]\int_{0}^{\infty}(\xi^2+1)^{d-2}d\xi=0,
\end{eqnarray}
where $\xi=\frac{x\pi}{m\alpha}$. As mentioned earlier, two types of Branch-cut terms, namely $\mathcal{B}_1(\alpha)$ and $\mathcal{B}_2(\alpha)$ appearing in eq.\,\eqref{after.Apsf.2.oddDim} are convergent. To obtain the value of $\mathcal{B}_1(a)$  we have:
\begin{eqnarray}\label{B1(a)-OddDim}
       \mathcal{B}_1(a)=i\int_{0}^{\infty}
       \frac{((it\pi)^2+m^2a^2)^{\frac{d-2}{2}}-((-it\pi)^2+m^2a^2)^{\frac{d-2}{2}}}
       {e^{2\pi t}-1}dt
       =\frac{2(-1)^{\frac{d-1}{2}}(ma)^{d-1}}{\pi}\int_{1}^{\infty}\frac{(\eta^2-1)^{\frac{d-2}{2}}}{e^{2ma\eta}-1}d\eta.
\end{eqnarray}
After expanding the denominator of the integrand and calculating the integrals, the final expression of $\mathcal{B}_1(a)$ becomes:
\begin{eqnarray}\label{B1(a)--Final form--Odd Dim}
       \mathcal{B}_1(a)=\frac{2(-1)^{\frac{d-1}{2}}(ma)^{d-1}}{\pi\sqrt{\pi}}\Gamma(d/2)\sum_{j=1}^{\infty}\frac{K_{\frac{d-1}{2}}(2maj)}{(maj)^{\frac{d-1}{2}}},
\end{eqnarray}
where $K_\nu(\alpha)$ is the modified Bessel function. For $\mathcal{B}_2(a)$ we obtain:
\begin{eqnarray}\label{B2(a)-OddDim}
        \mathcal{B}_2(a)&=&i\int_{0}^{\infty}
       \frac{((it\pi)^2+m^2a^2)^{d-2}-((-it\pi)^2+m^2a^2)^{d-2}}
       {e^{2\pi t}-1}dt=\frac{-2(ma)^{2d-3}}{\pi}\sin(d\pi)\int_{1}^{\infty}\frac{(\eta^2-1)^{d-2}}{e^{2ma\eta}-1}d\eta=0.\nonumber\\
\end{eqnarray}
As indicated in eq.\,\eqref{BSS.Def.}, the final step in the calculation of the Casimir energy is to compute the limit $L/b\to\infty$ and $b/a\to\infty$. Following the application of these limits, all contributions related to regions $A2$, $B1$, and $B2$ in eq.\,\eqref{after.Apsf.2.oddDim} are eliminated and the final expression of the total Casimir energy in odd spatial dimensions\,($d\neq1$) becomes:
\begin{eqnarray}\label{Final.Cas.Energy.OddDim}
      E^{(1)}_{\mbox{\tiny Cas.,Odd}}(a)=\frac{-\lambda L^{d-1}\pi^2\Omega_d^2}{32(2\pi)^{2d}a^{2d-3}}
     \left[\mathcal{B}_1(a)+\frac{2(ma)^{d-1}}{\pi}\left(\mathcal{H}(d)+\frac{(d-2)!!}{2(d-1)!!}\ln4\right)-(ma)^{d-2}\right]
     \mathcal{B}_1(a).
\end{eqnarray}
\begin{figure}[th] \hspace{-0.2cm}\includegraphics[width=8.5cm]{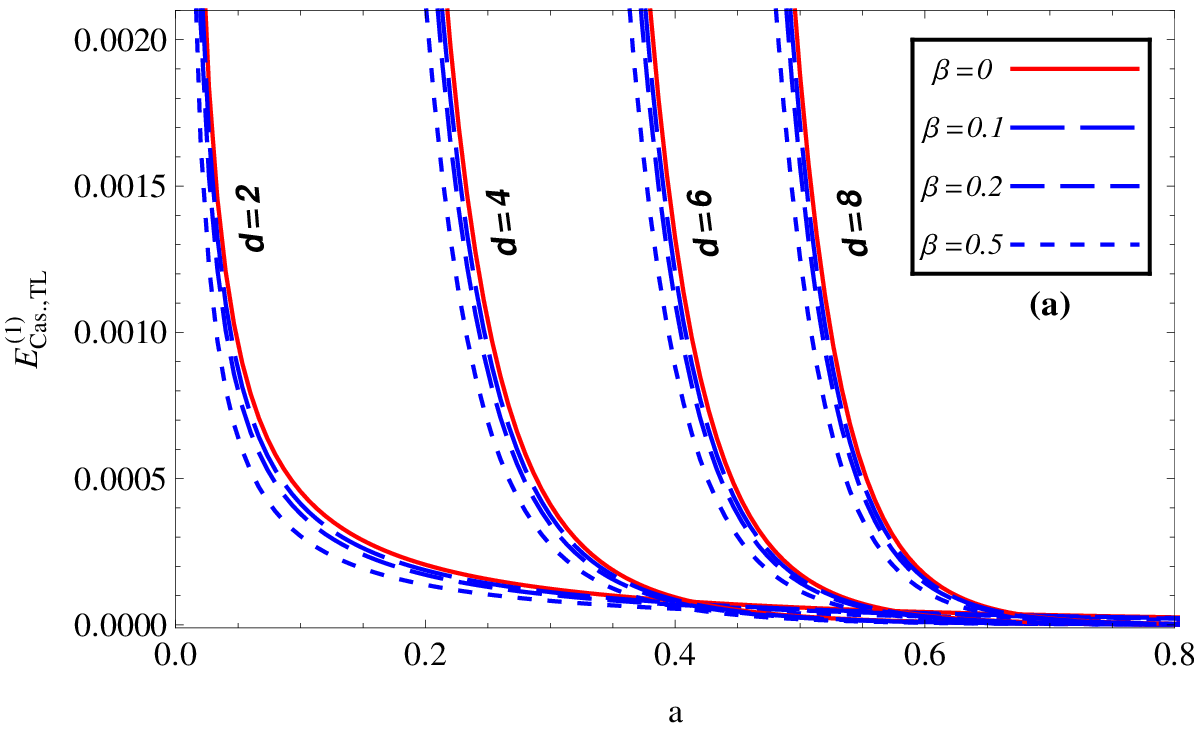}\hspace{0.5cm}\includegraphics[width=8.5cm]{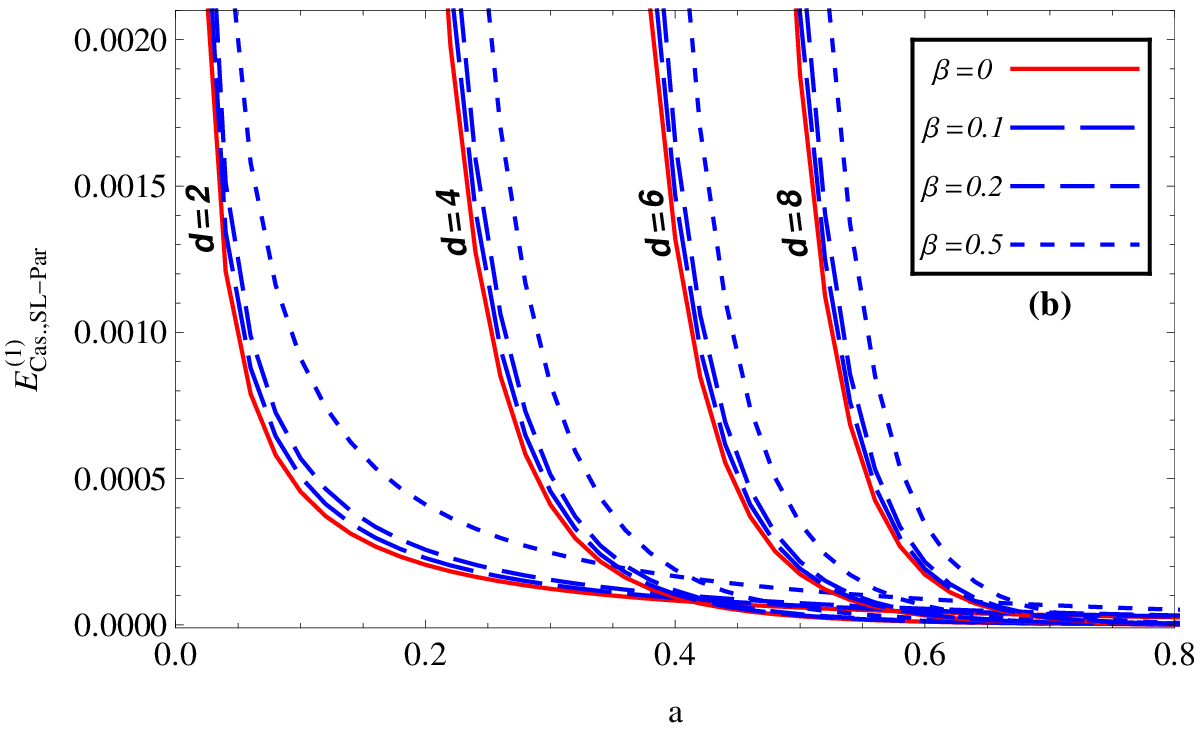}
\hspace{-2cm}\includegraphics[width=8.5cm]{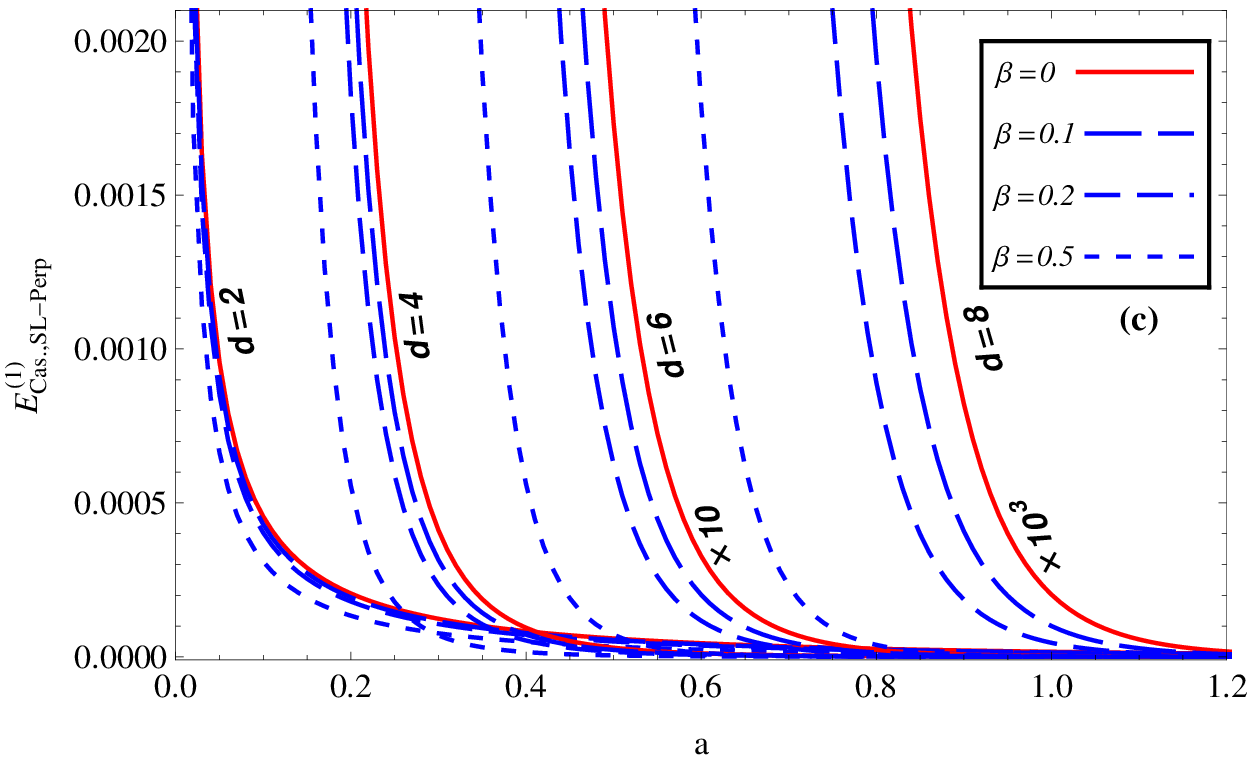}\hspace{0.5cm}\includegraphics[width=8.5cm]{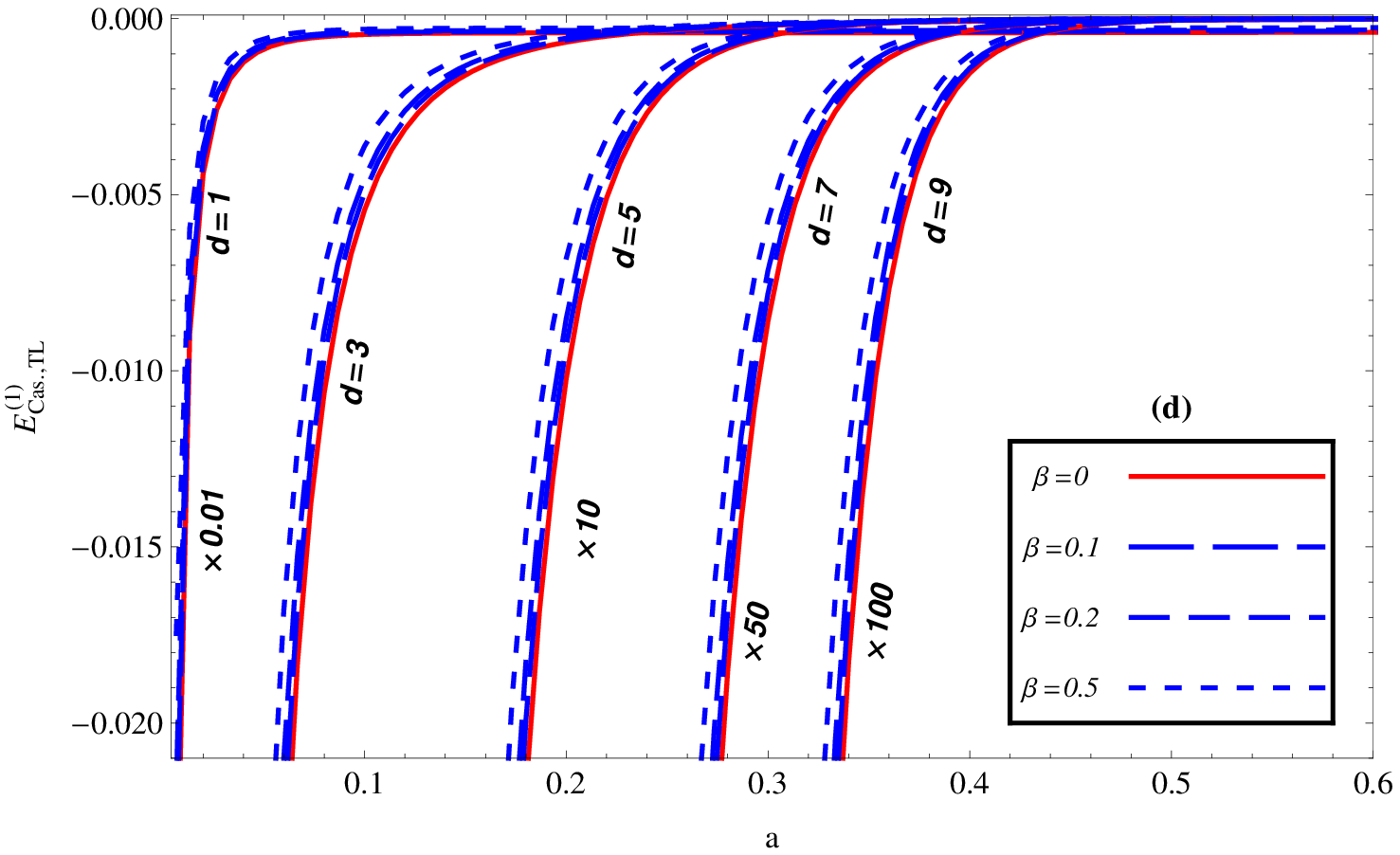}
\hspace{-2cm}\includegraphics[width=8.5cm]{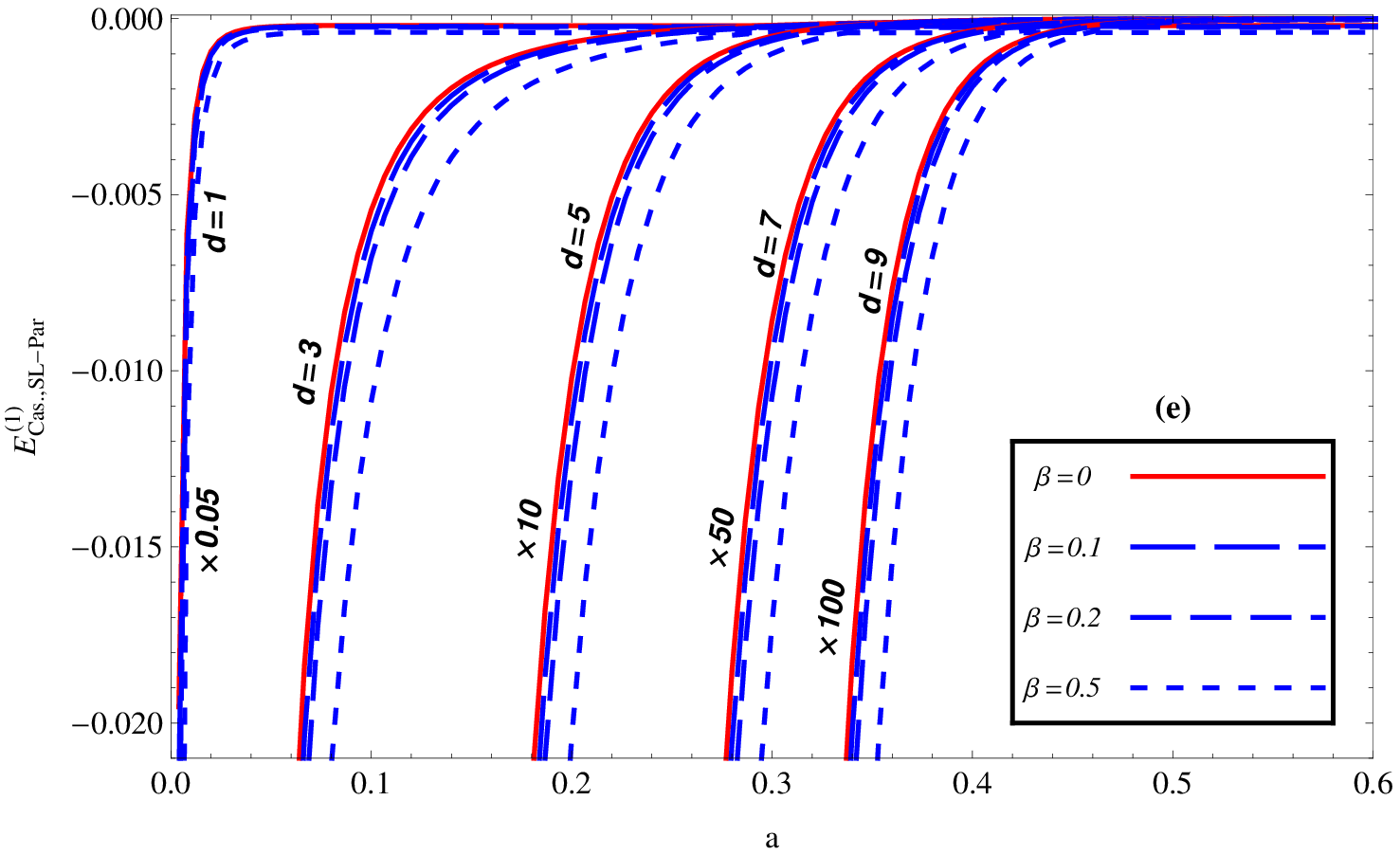}\hspace{0.5cm}\includegraphics[width=8.5cm]{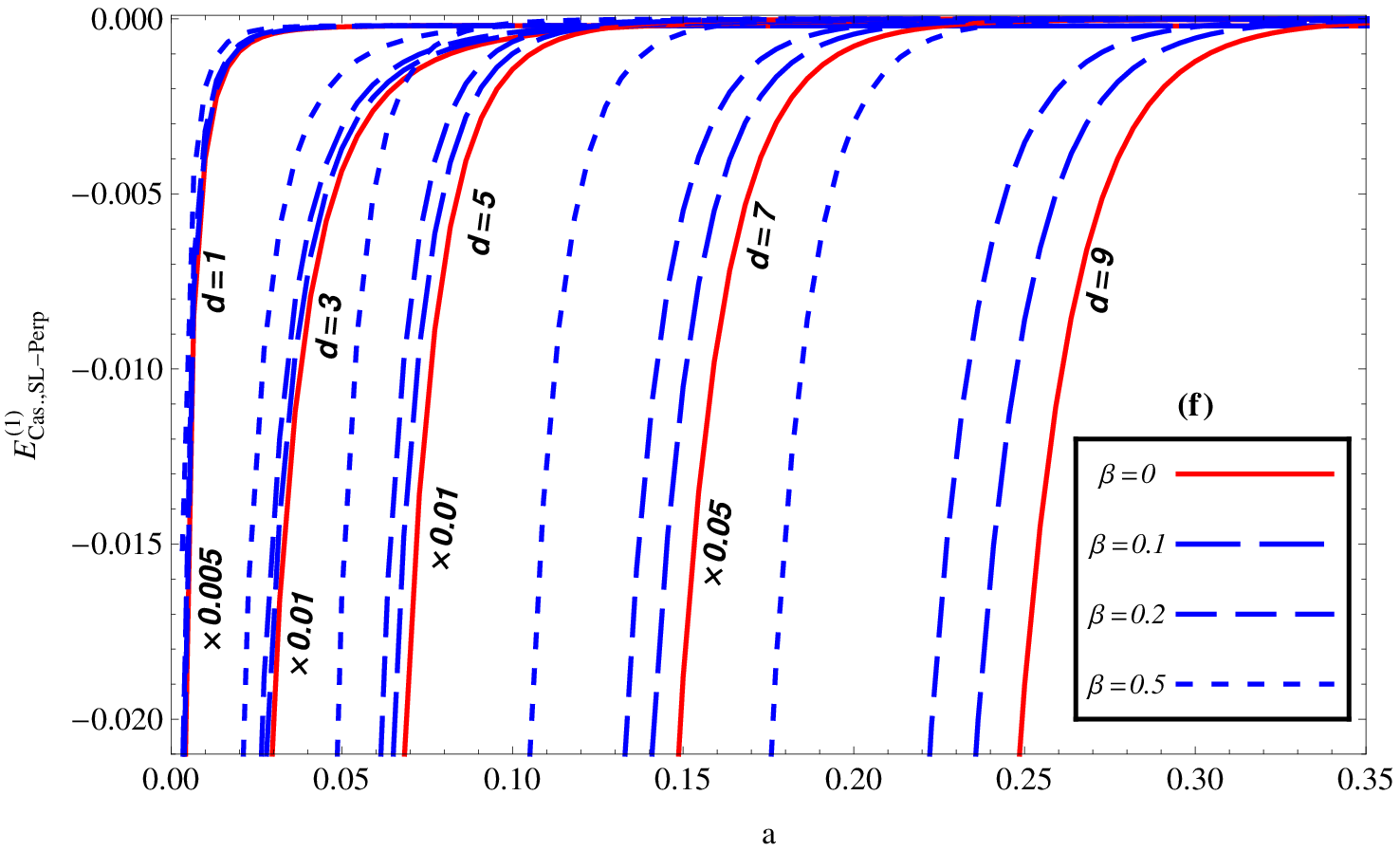}
\caption{\label{fig.LB.Odd&Even.TL&SLPA&SLPER} \small
  Figures (a), (b) and (c) plot the values of the first-order radiative correction to the Casimir energy density for massive and Lorentz-violating scalar field as a function of the distance of parallel plates\,($a$) in even spatial dimensions $d=\{2,4,6,8\}$. In figures (d), (e) and (f) this quantity is plotted in regard to odd spatial dimensions $d=\{1,3,5,7,9\}$. All plots were separated concerning the sequence values of $\beta=\{0,0.1,0.2,0.5\}$ in three violated directions of the Lorentz symmetry\,(TL, SL-Par and SL-Perp). In all plots, the value of coupling constant is $\lambda=0.1$.}
\end{figure}
An important extreme limit for the Casimir energy obtained in eq.\,\eqref{Final.Cas.Energy.OddDim} is the massless limit. To obtain the first-order radiative correction to the Casimir energy for the massless scalar field, we set the mass parameter $m=0$ in eq.\,\eqref{computing.the.integral.A1.ODD}. Therefore, we obtain:
\begin{eqnarray}\label{computing.the.integral.A1.Odd.MasslessCase}
E_{\mbox{\tiny Vac.}}^{(1)}(a)=\frac{-\lambda L^{d-1}\pi^2\Omega_{d}^2}{32(2\pi)^{2d}a^{2d-3}}\sum_{n,n'=1}^{\infty}
                                        \big(1+\frac{1}{2}\delta_{n,n'}\big)(n\pi)^{d-2}(n'\pi)^{d-2}, &\hspace{1.5cm}  d=3,5,7,9,....
\end{eqnarray}
By use of the analytic continuation technique in calculating the above summation, the radiative correction to the Casimir energy for the massless scalar field between two parallel plates in any odd spatial dimensions was obtained as\,($d\neq1$):
\begin{eqnarray}\label{Final.Cas.Energy.OddDim.Massless}
E^{(1)}_{\mbox{\tiny Cas.,Odd}}(m=0,a)=\frac{-\lambda L^{d-1}\Omega_d^2}{256\pi^2 (2a)^{2d-3}}
\Bigg[\zeta(2-d)^2+\frac{1}{2}\zeta(4-2d)\Bigg],
\end{eqnarray}
where $\zeta(\alpha)$ is the Riemann zeta function. Fig.\,(\ref{fig.ratio.Odd&Even}) plots the ratio of the Casimir energy correction values in two massive and massless cases for $d=\{3,5,7,9\}$. This figure satisfies the physical expectations and indicates a good consistency between the results obtained from eqs.\,\eqref{Final.Cas.Energy.OddDim} and \eqref{Final.Cas.Energy.OddDim.Massless}. Using position-dependent counterterms, \cite{3D-Reza} reported the radiative correction to the Casimir energy between two parallel plates for massive and massless scalar fields confined with Dirichlet boundary condition in $d=3$. Our obtained answers in eq.\,\eqref{Final.Cas.Energy.OddDim} and \eqref{Final.Cas.Energy.OddDim.Massless} regarding the specific case\,($d=3$) are completely consistent with the aforementioned study. Our results, on the other hand, differ from those reported in\,\cite{cavalcanti.1,cavalcanti.2,cavalcanti.3}. We maintain that the main source of this difference is that the counterterms used in the present research were automatically extracted from the $n$-point function and were consistent with the imposed boundary condition. However, the counterterms in the renormalization program employed in the reported works, regardless of the type of the imposed boundary conditions, were free counterterms. To obtain the radiative correction to the Casimir energy for Lorentz-violating scalar field in any odd spatial dimensions\,($d\neq1$), the Green's function expressions given in eqs.\,\eqref{Green.function.TLDBC}, \eqref{Green.function.SLPADBC} and \eqref{Green.function.SLPRDBC} should be substituted with the vacuum energy expression displayed in eq.\,\eqref{VacuumEn.firstorder.}. Next, all computation processes should be repeated in the same way as what occurred in this subsection for the odd spatial dimensions. Doing so, the radiative correction to the Casimir energy for massive and Lorentz-violating scalar field in odd spatial dimensions is:
\begin{eqnarray}\label{Final.Cas.Energy.OddDim.LB.TL&SLPA.&SLPR.Massive}
     E^{(1)}_{\mbox{\tiny Cas.,TL}}(a)&=&\frac{1}{1+\beta}E^{(1)}_{\mbox{\tiny Cas.,Odd}}(a),\nonumber\\
     E^{(1)}_{\mbox{\tiny Cas.,SL-Par}}(a)&=&\frac{1}{1-\beta}E^{(1)}_{\mbox{\tiny Cas.,Odd}}(a),\nonumber\\
     E^{(1)}_{\mbox{\tiny Cas.,SL-Perp}}(a)&=&E^{(1)}_{\mbox{\tiny Cas.,Odd}}(\tilde{a}).
\end{eqnarray}
where $\tilde{a}=\frac{a}{\sqrt{1-\beta}}$. To achieve the radiative correction to the Casimir energy for the massless and Lorentz-violating scalar field, we set the mass parameter $m=0$ in eqs.\,\eqref{Green.function.TLDBC}, \eqref{Green.function.SLPADBC} and \eqref{Green.function.SLPRDBC}. To obtain the vacuum energy in each violated direction of the Lorentz symmetry breaking, the related Green's function was then substituted in eq.\,\eqref{VacuumEn.firstorder.}. Ultimately, by applying the analytic continuation technique to the vacuum energy expression, we obtain the radiative correction to the Casimir energy for the massless scalar field in the Lorentz-violating system as:
\begin{eqnarray}\label{Final.Cas.Energy.OddDim.LB.TL&SLPA.&SLPR.Massless}
     E^{(1)}_{\mbox{\tiny Cas.,TL}}(m=0,a)&=&\frac{1}{1+\beta}E^{(1)}_{\mbox{\tiny Cas.,Odd}}(m=0,a),\nonumber\\
     E^{(1)}_{\mbox{\tiny Cas.,SL-Par}}(m=0,a)&=&\frac{1}{1-\beta}E^{(1)}_{\mbox{\tiny Cas.,Odd}}(m=0,a),\nonumber\\
     E^{(1)}_{\mbox{\tiny Cas.,SL-Perp}}(m=0,a)&=&E^{(1)}_{\mbox{\tiny Cas.,Odd}}(m=0,\tilde{a}).
\end{eqnarray}
Figs.\,(\ref{fig.LB.Odd&Even.TL&SLPA&SLPER}d), (\ref{fig.LB.Odd&Even.TL&SLPA&SLPER}e), and (\ref{fig.LB.Odd&Even.TL&SLPA&SLPER}f) plot the radiative correction to the Casimir energy values as a function of the distance of plates\,($a$) for $d=\{3,5,7,9\}$ and $\beta=\{0,0.1,0.2,0.5\}$. This sequence of plots is shown in separate figures regarding three violated directions of the Lorentz symmetry breaking\,(TL, SL-Par and SL-Perp). Fig.\,(\ref{fig.LB.Odd&Even.TL&SLPA&SLPER}) shows that the first-order correction to the Casimir energy for all odd spatial dimensions was negative, and the effect of Lorentz symmetry breaking on the Casimir energy value in the case of SL-Par was higher than the other violated directions of the Lorentz symmetry breaking. Fig.\,(\ref{fig.LB.Odd&Even.TL&SLPA&SLPER}) also shows the Lorentz violating in the time-like case had the minimum effect on the value of the Casimir energy.
\par
\textcolor[rgb]{0.00,0.00,0.00}{The interesting point in this problem is whether the pure contribution arising from the Lorentz symmetry breaking in the leading-order of the Casimir energy can cancel the radiative correction term of the Casimir energy in the system holding the Lorentz symmetry. To address this point, as a simple example, we started with the leading-order of the Casimir energy regarding the massless and TL Lorentz violated scalar field obeying Dirichlet boundary conditions on a pair of plates within three spatial dimensions:
\begin{eqnarray}\label{Leading.TL.Cas.3D}
      E^{(0)}_{\mbox{\tiny Cas.,TL}}(a)=-\frac{\sqrt{1+\beta}L^2\pi^2}{1440a^3},
\end{eqnarray}
where superscript $(0)$ indicates the leading-order of the Casimir energy. Expanding the expression of eq.\,(\ref{Leading.TL.Cas.3D}) in the limit $\beta\to0$, we simply obtain:
\begin{eqnarray}\label{Leading.TL.Cas.3D.expanding .beta}
     E^{(0)}_{\mbox{\tiny Cas.,TL}}(a)=-\frac{L^2\pi^2}{1440a^3}-\frac{\beta L^2\pi^2}{2880a^3}+\mathcal{O}(\beta)^2,
\end{eqnarray}
Furthermore, using eq.\,(\ref{Final.Cas.Energy.OddDim.LB.TL&SLPA.&SLPR.Massless}), the expression of radiative correction to the Casimir without Lorentz violation is obtained as:
\begin{eqnarray}\label{RC.3D.Massless.withoutLV}
E^{(1)}_{\mbox{\tiny Cas.,Odd}}(m=0,a)=\frac{-\lambda L^2}{18432a^3}.
\end{eqnarray}
Comparison of eq.\,(\ref{RC.3D.Massless.withoutLV}) to the second term on the right hand side of eq.\,(\ref{Leading.TL.Cas.3D.expanding .beta}) shows that, by choosing $\beta=\frac{-5\lambda}{32\pi^2}$, the pure contribution arising from the TL Lorentz symmetry breaking in the leading-order Casimir energy can cancel the radiative correction term in the system without the Lorentz violation. Adjustment of the same value for the parameter $\beta$ in SP-Par like Lorentz symmetry breaking can remove the radiative correction term. Concerning SP-Perp Lorentz violation, this cancellation process will occur by $\beta=\frac{5\lambda}{96\pi^2}$. Performing the aforementioned process for the massive scalar field makes that the relation between parameter $\beta$ and coupling constant $\lambda$ would be obtained as mass-dependent. For instance, in three spatial dimensions for the massive and TL Lorentz violation scalar field with the mass of $m=1$, to cancel the radiative correction term with the pure Lorentz violating contribution, the value of parameter was obtained as $\beta\approx-0.0165\lambda$. This value of $\beta$ alters with changing the mass value. We maintain that this process can be generalized in other spatial dimensions.}
\section{Conclusion}\label{sec:conclusion}
In this study, we computed the radiative correction to the Casimir energy for the Lorentz-violated massive and massless scalar field confined with Dirichlet boundary condition between a pair of parallel plates in $d$ spatial dimensions. The main prominent point in this calculation was the type of counterterm used in the renormalization program. \cite{cavalcanti.1,cavalcanti.2,cavalcanti.3} previously reported the Dirichlet Casimir energy between two parallel plates in every spatial dimension for the system where the Lorentz symmetry was preserved. However, the results obtained in these works were reported divergent for all even dimensions. We maintain that the main source of this divergence is attributed to the type of the employed counterterm. The position-dependent counterterm allows all effects of the boundary conditions or non-trivial backgrounds to be imported in the renormalization program. This type of counterterm also creates a self-consistent manner to renormalize the bare parameters of the Lagrangian in the renormalization program. In the present study, through the use of the position-dependent counterterm, we computed the radiative correction to the Casimir energy for massive and massless scalar field between two parallel plates in all spatial dimensions. This calculation was further generalized to the Lorentz violating scalar field. Our final answers for the radiative correction to the Casimir energy were convergent for all spatial dimensions and consistent with the expected physical grounds. In all spatial dimensions, the Casimir energies of the massive and massless scalar fields approached each other in the appropriate limits. The problem considered in this work was solved for a number of spatial dimensions\,(\emph{e.g.}, $d=1,2$ and $3$) using the position-dependent counterterms in\,\cite{1D-Reza,2D-Man,3D-Reza}. Our result is a generalization of those works regarding all spatial dimensions in Lorentz-violating system and is in line with their results in the appropriate limits\,(\emph{e.g.}, $d=1,2$ and $3$). \textcolor[rgb]{0.00,0.00,0.00}{In three spatial dimensions, we demonstrate that the pure contribution arising from the Lorentz symmetry breaking in the leading-order Casimir energy can cancel the radiative correction term\,($\mathcal{O}(\lambda)$) in the system without the Lorentz violation. This issue was demonstrated for massive and massless scalar fields within three spatial dimensions, and it may be generalizable to other space-time dimensions.}

\appendix\section{Radiative correction to the Casimir energy for Lorentz-violating scalar field in one spatial dimension}\label{appendix.1Dim}
To obtain the radiative correction to the Casimir energy for the massive scalar field confined with Dirichlet boundary condition between two points with distance $a$ in $1+1$ dimensions, we commence with the related Green's function expression given in eq.\,\eqref{Greens.Function}. We set $d=1$ in this equation and after substituting the Green's function in eq.\,\eqref{VacuumEn.firstorder.}, the vacuum energy is obtained as:
\begin{eqnarray}\label{vacuum.En.1Dim}
    E_{\mbox{\tiny Vac.}}^{(1)}(a)=\frac{\lambda\pi^2}{8a}
    \bigg[\Big(\sum_{n=1}^{\infty}\frac{1}{\sqrt{\frac{n^2\pi^2}{a^2}+m^2}}\Big)^2+\frac{ma\coth ma-1}{4m^2}\bigg].
\end{eqnarray}
To regularize this infinite expression and obtain the Casimir energy, the BSS given in eq.\,\eqref{BSS.Def.} should be employed. The details of the calculation were previously reported in\,\cite{1D-Reza}. Therefore, we do not repeat the calculation here and only report the final answer of the radiative correction to the Casimir energy as:
\begin{eqnarray}\label{Cas.En.TL.1Dim}
   E_{\mbox{\tiny Cas.}}^{(1)}(d=1;m,a)=\frac{-\lambda\pi^2}{8}
   \bigg[B(a)\Big(\frac{2\ln2}{\pi}+\frac{B(a)}{a}-\frac{1}{ma}\Big)+\frac{\coth(ma)}{4m}\bigg],
\end{eqnarray}
where $B(a)$ is:
\begin{eqnarray}\label{Branchcut.1Dim}
    B(a)=\frac{2a}{\pi}\int_{1}^{\infty}\frac{(\eta^2-1)^{-1/2}}{e^{2ma\eta}-1}d\eta.
\end{eqnarray}
This result for the radiative correction to the Casimir energy was obtained in the system where the Lorentz symmetry was still preserved\,($\beta=0$). \textcolor[rgb]{0.00,0.00,0.00}{In ($1+1$)-dimensions there exist only one space-like unit vector $u^\mu=(0,1)$.} Therefore, to obtain the vacuum energy for the Lorentz violating system, we begin with the related Green's function for each Lorentz violated direction given in eqs.\,\eqref{Green.function.TLDBC} and \eqref{Green.function.SLPRDBC}. Then, we set the spatial dimension $d=1$. The obtained Green's function expression should be substituted with the vacuum expression written in eq.\eqref{VacuumEn.firstorder.}. Therefore, we obtain,
\textcolor[rgb]{0.00,0.00,0.00}{\begin{eqnarray}\label{Vac.En.TL&SLPar&SLPerp.1Dim}
    E_{\mbox {\tiny Vac.,TL}}^{(1)}(a)&=&\frac{1}{1+\beta} E_{\mbox {\tiny Vac.}}^{(1)}(a),\nonumber\\
    E_{\mbox {\tiny Vac.,SL-Perp}}^{(1)}(a)&=& E_{\mbox {\tiny Vac.}}^{(1)}(\tilde{a}),
\end{eqnarray}}
where $\tilde{a}=\frac{a}{\sqrt{1-\beta}}$. All steps of achieving the radiative correction to the Casimir energy from each of the above expressions are similar to those previously conducted to obtain the Casimir energy for a system with Lorentz symmetry preservation. Therefore, there is no need to repeat those steps, and according to the vacuum energy expressions written in eq.\,\eqref{Vac.En.TL&SLPar&SLPerp.1Dim}, the Casimir energy for each Lorentz violating directions can be obtained as follows:
\textcolor[rgb]{0.00,0.00,0.00}{\begin{eqnarray}\label{Cas.En.TL&SLPar&SLPerp.1Dim}
    E_{\mbox {\tiny Cas.,TL}}^{(1)}(a)&=&\frac{1}{1+\beta} E_{\mbox {\tiny Cas.}}^{(1)}(d=1;m,a),\nonumber\\
    E_{\mbox {\tiny Cas.,SL-Perp}}^{(1)}(a)&=& E_{\mbox {\tiny Cas.}}^{(1)}(d=1;m,\tilde{a}).
\end{eqnarray}}
The radiative correction to the Casimir energy for the massless scalar field was reported zero\,\cite{1D-Reza}. Accordingly, given the relationships between the Green's function expressions concerning each violated direction of the Lorentz symmetry written in equations\,\eqref{Green.function.TLDBC} and \eqref{Green.function.SLPRDBC}, it can simply be said that the value of radiative correction to the Casimir energy regarding the massless and Lorentz-violated scalar fields in $1+1$ dimensions is also zero.

\acknowledgments
\textcolor[rgb]{0.00,0.00,0.00}{The author would also like to thank the reviewers of MPLA journal for their meticulous review of the article and highly useful suggestions that improved the quality of the article. Additionally, the Author would like to thank the research office of, Islamic Azad University, Semnan Branch for its financial support. }

\end{document}